\documentclass[journal]{vgtc}                % final (journal style)

\ifpdf%                                % if we use pdflatex
  \pdfoutput=1\relax                   % create PDFs from pdfLaTeX
  \usepackage{graphicx}                % allow us to embed graphics files
  \DeclareGraphicsExtensions{.pdf,.png,.jpg,.jpeg,.PNG,.JPG} % for pdflatex we expect .pdf, .png, or .jpg files
\else%                                 % else we use pure latex
  \usepackage{graphicx}                % allow us to embed graphics files
  \DeclareGraphicsExtensions{.eps}     % for pure latex we expect eps files
\fi%

\graphicspath{{}{pictures/}{images/}{./}} % where to search for the images

\usepackage{microtype}                 % use micro-typography (slightly more compact, better to read)
\PassOptionsToPackage{warn}{textcomp}  % to address font issues with \textrightarrow
\usepackage{textcomp}                  % use better special symbols
\usepackage{mathptmx}                  % use matching math font
\usepackage{times}                     % we use Times as the main font
\usepackage{cite}                      % needed to automatically sort the references
\usepackage{tabu}                      % only used for the table example
\usepackage{booktabs}                  % only used for the table example
\onlineid{6}

\vgtccategory{Research}
\vgtcpapertype{Technique}

\usepackage{hyperref}
\usepackage{booktabs} % For formal tables

\usepackage{epstopdf}% To incorporate .eps illustrations using PDFLaTeX, etc.
\usepackage{subfigure}% Support for small, `sub' figures and tables
\usepackage{microtype}                 % use micro-typography (slightly more compact, better to read)
\PassOptionsToPackage{warn}{textcomp}  % to address font issues with \textrightarrow
\usepackage{textcomp}                  % use better special symbols
\usepackage{mathptmx}                  % use matching math font
\usepackage{times}                     % we use Times as the main font
\usepackage{cite}                      % needed to automatically sort the references
\usepackage{tabu}                      % only used for the table example
\usepackage{booktabs} 
\usepackage{ragged2e}% only used for the table example
\usepackage[dvipsnames]{xcolor}
\usepackage{csquotes}
\usepackage{fontawesome}
\usepackage{amsmath}
\usepackage{amssymb}
\usepackage{graphicx}
\usepackage{calc}
\usepackage{tikz}
\usepackage[export]{adjustbox}
\usepackage{wrapfig}
\usepackage{cancel}
\usepackage{multirow}
\usepackage{fontawesome}
\usepackage{xcolor}
\usepackage{caption}
\usepackage{soul}
\usepackage{microtype} 

\definecolor{light-gray}{gray}{0.95}

\definecolor{col-topic}{RGB}{38,130,156}
\definecolor{col-concept}{RGB}{98,139,45}
\newcommand{\topic}[1]{\textcolor{col-topic}{#1}}
\newcommand{\concept}[1]{\textcolor{col-concept}{#1}}

\newcommand{\revision}[1]{#1}

\newlength\myheight
\newlength\mydepth
\settototalheight\myheight{Xygp}
\newcommand*\inlinegraphics[1]{%
  \settototalheight\myheight{Xygp}%
  \settodepth\mydepth{Xygp}%
  \raisebox{-\mydepth}{\includegraphics[height=\myheight]{#1}}%
}

\title{Semantic Concept Spaces: Guided Topic Model Refinement\\ using Word-Embedding Projections}
\author{
Mennatallah El-Assady$^{1,2}$, Rebecca Kehlbeck$^1$, %\\ % Miriam Butt,
	 Christopher Collins$^2$, Daniel Keim$^1$, and Oliver Deussen$^1$ \\[0.5em]
	 $^1$University of Konstanz, Germany.
\qquad $^2$Ontario Tech University, Canada.}

\abstract{
We present a framework that allows users to incorporate the semantics of their domain knowledge for topic model refinement while remaining model-agnostic. Our approach enables users to (1)~\textit{understand} the semantic space of the model, (2)~\textit{identify} regions of potential conflicts and problems, and (3)~\textit{readjust} the semantic relation of concepts based on their understanding, directly influencing the topic modeling. These tasks are supported by an interactive visual analytics workspace that uses word-embedding projections to define \textit{concept regions} which can then be refined. The user-refined concepts are independent of a particular document collection and can be transferred to related corpora.  All user interactions within the concept space directly affect the semantic relations of the underlying vector space model, which, in turn, change the topic modeling.  
In addition to direct manipulation, our system guides the users' decision-making process through recommended interactions that point out potential improvements.  This targeted refinement aims at minimizing the feedback required for an efficient human-in-the-loop process. We confirm the improvements achieved through our approach in two  
user 
studies that show topic model quality improvements through our visual knowledge externalization and learning process.
}

\shortauthortitle{El-Assady \MakeLowercase{\textit{et al.}}: Semantic Concept Spaces}

\CCScatlist{ % not used in journal version
 \CCScat{K.6.1}{Management of Computing and Information Systems}%
{Project and People Management}{Life Cycle};
 \CCScat{K.7.m}{The Computing Profession}{Miscellaneous}{Ethics}
}

\teaser{
\centering
  \includegraphics[width=0.928\textwidth]{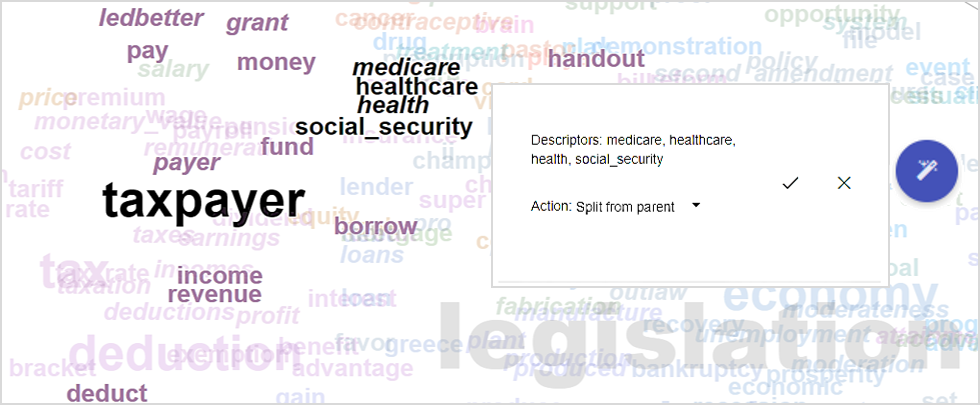}
  \caption{Guided relevance feedback for the targeted refinement of incoherent areas in the \textit{Semantic Concept Space}. This user guidance component tours through the space and highlights potentially uncertain areas, suggesting a recommended action for refinement.  
  }
\label{fig:teaser}
}

\begin{document}
\firstsection{Introduction}
\maketitle
 Efficiently categorizing  the contents of large text collections into thematic groups is a common task for  scholars in the humanities and social sciences. These data and domain experts usually embark on a process of summarizing documents, extracting concepts, modeling their relations, and finally, aggregating the obtained information to build their knowledge. 
The generated knowledge is typically externalized in various resources, including traditional books and papers, but also extensive knowledge bases~\cite{sadeghi2017integration}. However, even given the eagerness with which experts strive to model and document their knowledge and intuition, oftentimes available resources do not capture all specific aspects of a domain's semantics~\cite{federico2017role}. 
The shortage of domain-specific knowledge representations in accessible formats has sparked a bustling research area~\cite{wang2014knowledge} at the intersection of linguistics and machine learning.

% On the other hand, automatic, data-driven techniques, such as topic modeling, are becoming more reliable and accessible. 
Simultaneously, domain-knowledge-independent machine learning techniques are becoming more reliable and accessible. 
For instance, topic modeling algorithms have wide applicability across a multitude of domains as they augment the time-consuming task of categorizing document collections into thematically-related groups. 
Despite their usefulness, the quality of their results highly depends on the suitability of the parameter choices and how well they fit and reflect the characteristics of the analyzed document collection and domain semantics. 
However, as such models are typically black boxes, they are not readily understood by non-machine-learning-experts.   
Thus, there is a need for %interactive 
machine learning refinement techniques that abstract the complexity of underlying models, enabling users to \textit{understand}, \textit{diagnose}, and \textit{refine} the results. This user demographic does not desire to understand the inner-workings of machine learning but would rather to \textit{teach} the machine their semantic knowledge while remaining \textit{model-agnostic}.

Promising visual analytics solutions have been proposed to address such challenges in a collaborative human-machine effort. For example, to model the semantic relations of concepts in a corpus,   \textit{ConceptVector}~\cite{Park2017ConceptVector:Embedding} has been proposed as an interactive lexicon building approach using word embeddings. On the other hand, \textit{UTOPIAN}~\cite{Choo2013} enables users to interactively train a topic model, resulting in a clustering of documents into thematic groups. While the first approach is designed to consider the user's knowledge for \textit{top-down} concept generation, the second one is data-driven, generating topics \textit{bottom-up}. Techniques combining high-level analysis concepts with low-level model interaction, e.g. through \textit{bidirectional} semantic interaction~\cite{dowling2018bidirectional}, have proven effective since \enquote{\textit{the power of the computational models can be leveraged without their complexity}}~\cite{endert2015semantic}.

\revision{We present a visual analytics technique that tightly links these two perspectives to allow users to externalize their domain knowledge for topic model refinement without understanding the inner-workings of such models.} \revision{Our lead motivation for such an iterative refinement}
\revision{process} is to enable users to \textit{teach}~\cite{simard2017machine} the machine learning model
\revision{(through \textit{concept refinement})}, and in turn, the model to respond by \textit{learning} a new refined representation \revision{(through a \textit{topic model update})} that is presented to the users to show them the effects of their interactions.
Hence, our technique relies on two independent hierarchical 
structures, (1)~the \textbf{concept hierarchy}, representing the user's semantics \revision{(\textit{top-down})}, and (2)~the \textbf{topic hierarchy} that is based on the automatically computed results of a topic model \revision{(\textit{bottom-up})}. \revision{Both hierarchies operate on the same vector space but are presented in two separate views. %They are two sides of the same coin. 
The \textit{concept view} is used as the \textit{interactive} view for domain knowledge externalization, while the \textit{topic view} is a \textit{reactive} component for inspecting the topic model updates caused by refining semantic relations in the first view.}  
This \textit{duality} is captured in the topic and concept representations as two superimposed canvases, facilitating the analysis of associations~\cite{Gleicher2018ConsiderationsComparison}.

Thus, the main challenge for our technique is to define accurate mappings from the users' interactions back to \textit{actionable} instructions for the topic model optimization. On the visualization side, %another 
the challenge is to find an accurate and faithful \textit{spatialization} of concepts and topics on a canvas, while reducing clutter and retaining semantic neighborhoods. 

We designed \textit{Semantic Concept Spaces} as a mixed-initiative technique tailored to support users in modeling their domain knowledge through defining semantic relations between concepts. 
Our approach (1)~provides different entry points and abstraction layers for the users' analysis; (2)~integrates users in every step of the semantic concept creation; (3)~allows for \textit{targeted refinement} through guided relevance feedback, as well as, \textit{concept discovery} through serendipitous exploration; (4)~enables cross-corpus and model-agnostic learning to allow the transferability of the learned concepts to other topic models and similar document collections; and (5)~abstracts from the refined semantics to update domain-specific concepts, avoiding future \textit{cold starts}~\cite{Schein2002MethodsRecommendations}.

\begin{figure}
	\centering
	\includegraphics[width=\columnwidth]{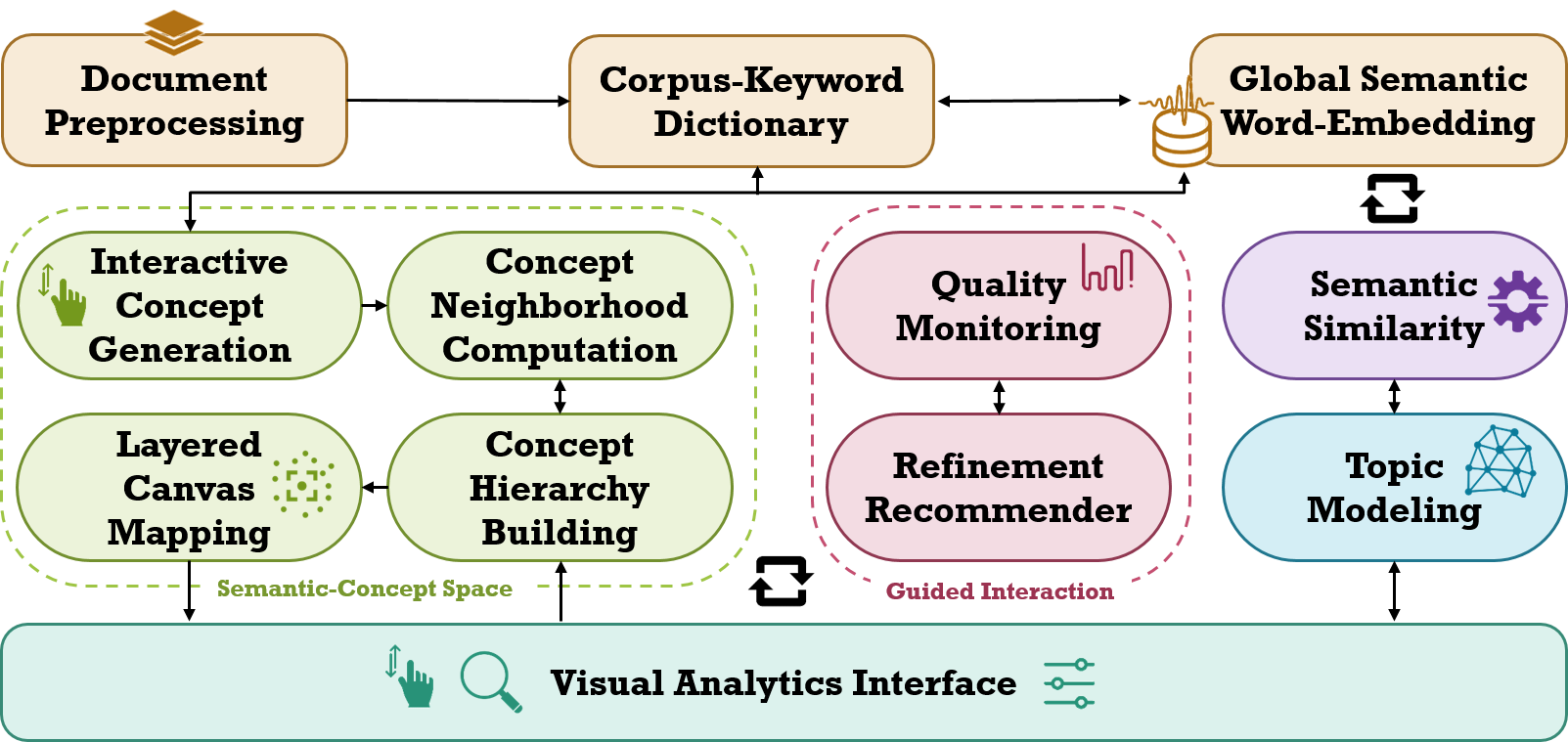}
	\caption{The human-in-the-loop workflow for Semantic Concept Spaces. }
	\label{fig:Pipeline}
	\vspace{-15pt}
\end{figure}

\autoref{fig:Pipeline} depicts the architecture of our approach, starting with processing a document collection to extract relevant keywords and embeddings~\cite{levy2014dependency}. These build the basis for the semantic similarity that generates scored keyword vectors as input for topic modeling, they also initialize the interactive concept generation. 
This step extracts seed-words for the concept \revision{generation}, optionally including user-defined structures. 
% To define the spatialization of the \revise{linguistic fabric} for the visualization, concept neighborhoods are calculated using t-SNE~\cite{hinton2003stochastic}.
\revision{To define a meaningful spatialization}, concept neighborhoods are calculated using t-SNE~\cite{hinton2003stochastic}.
After the building of the initial concept hierarchy, all elements of the visualization are projected~\cite{maaten2008visualizing} onto a canvas in layers. The visual analytics interface is the main workspace for the user's interaction, this enables users to inspect concepts and topics to \textbf{[T1]}~\textbf{understand} their relationships, \textbf{[T2]}~\textbf{diagnose} potential conflicts, \textbf{[T3]}~\textbf{refine} the concept space based on their domain understanding, \revision{ and \textbf{[T4]}~\textbf{update} the topic modeling based on the refined concept space}.  A continuous quality monitoring and refinement recommendation  \revision{supports these tasks and enables targeted user guidance.}

We evaluated our technique with three approaches. Starting with a mixed-method expert study, six participants used Semantic Concept Spaces on a model refinement task. Second, a quantitative evaluation of the model improvement achieved by experts, across eight model quality measures. Finally, four independent annotators rated the quality of these concept spaces and topic model results.

In summary, this work contributes an iterative visual analytics approach that captures user semantic knowledge to inform   machine learning systems. 
Our technique provides user guidance and relevance feedback for overcoming the ``looseness'' of the interaction mapping.  We demonstrate and test it on a case study in topic model refinement.

\section{Background and Related Work}

This research is an entry into the burgeoning space of interactive and explainable machine learning applications~\cite{endert2017state,Liu2017TowardsPerspective,jiang2018recent}. In the following discussion, we will relate our work to research in the areas of semantic interaction for visual analytics, and, more specifically, interactive topic modeling and content analysis. 

\paragraph{Semantic Interaction}

We are inspired by the call by Endert et al. for interaction ``beyond control panels," where data is spatialized and the interaction with that \textit{spatialization} is the primary mechanism for manipulating the data space~\cite{endert2013beyond}. Semantic interaction is a type of direct feedback users can provide to embed their semantic understanding into a visual workspace~\cite{Enderta,endert2015semantic}. In this paradigm, user interactions are used to feed-forward into model refinements. For example, the global layout in \textit{ForceSPIRE}~\cite{Enderta} is adjusted based on users moving words and documents to externalize their knowledge in the workspace. Cavallo and Demiralp~\cite{Cavallo2018} employ both forward and backward-projection interactions to enable user interaction with the dimension reduction algorithms. Forward-projection enables users to change the high dimensional vector and see how the projection is changed, and backwards-projection uses direct manipulation to move nodes and see how the input vector changes. In the present work we use semantic interaction to enable users to modify a \textit{word2vec} \revision{word} embedding space~\cite{Mikolov2013} by modifying the groupings of words into concepts. As there may be many ways to adjust the semantic space, we provide suggested interactions as a form of guidance~\cite{collins2018guidance}.

\paragraph{Topic Modeling and Content Analysis}

Topic Modeling is used to understand large corpora of text and summarize the knowledge contained in them. The basic premise of topic modeling is to cluster groups of documents and label them, obtaining topics. An overview of  probabilistic topic modeling algorithms can be seen in the survey by Blei~\cite{Blei2012ProbabilisticModels}. Several works explicitly address the embedding of the domain knowledge into the topic space. Andrzejewski et al.~\cite{Andrzejewski2009IncorporatingPriors} use Dirichlet forest priors to split and merge concepts using domain knowledge, improving topic descriptors.
Chen et al. developed the MDK-LDA  variant on LDA which takes into account domain knowledge directly to provide better topic descriptors~\cite{Chen2013LeveragingModels}. \revision{Furthermore, approaches that combine word embeddings with topic modeling can be beneficial for learning both models jointly~\cite{moody2016mixing}, as well as improving topic model representations for short texts through word embeddings~\cite{nguyen2015improving,zhao2017word,li2017enhancing}, or creating improved word embeddings using LDA~\cite{ren2016topic}.} 

Exploratory visualizations for understanding topic spaces include  \textit{ParallelTopics}~\cite{Dou2011ParallelTopics:Collections}, for exploring single and multi-topic documents using parallel coordinates.
The focus of our work, however, is not on viewing the topic modeling itself, but  finding an intuitive way for the users to guide the modeling process. Visual topic modelling approaches often include some interactive mechanisms for users to modify the modeling output.   \textit{TopicPanorama}~\cite{Wang2016TopicPanorama:Topics} creates a graph of topic relations extracted from multiple sources. Interactive tools are embedded to allow users to modify the graph matching to suit their needs.  \textit{HierarchicalTopics}~\cite{Dou2013HierarchicalTopics:Hierarchies} is a visualization for understanding a large dataset at different levels of granularity. Interactions allow users to adjust the hierarchy. \textit{UTOPIAN} employs a semi-supervised iterative feedback loop for users to steer the modeling process~\cite{Choo2013}. 
 \textit{ConceptVector}~\cite{Park2017ConceptVector:Embedding} enables users to embed domain knowledge interactively, through guiding the building of concepts which are then used to analyze documents. We follow a similar approach in allowing users to refine the concept space which is used as the substrate for topic modeling. This idea of an interactive loop for refining topics appears also in the work of Hu et al.~\cite{hu2014interactive}, in which users guide the modeling process through constraints on the topic descriptors. Hoque and Carenini embed similar feedback into a visualization system in the \textit{ConVisIt} project~\cite{hoque2015convisit}. In our previous work, we reported a user-guided refinement process for topic modeling based on ``voting'' for models which have subjectively higher quality~\cite{El-Assady2018ProgressiveFramework}. Speculative execution has also been used to preview the outputs of topic modeling and allow users to intervene and guide the process~\cite{el2018visual}.

\section{Modeling the Semantic Concept Space} 
\label{sec:modeling}

To model the \revision{semantic concept space} %\revise{linguistic fabric} 
of a corpus, we consider all the words it contains and all their embeddings as a foundation (these are a subset of all word in a language's vocabulary). Based on this set of words, we build two separate, parallel hierarchies; the concept and the topic hierarchies. Both contain four abstraction levels, sharing the lowest level of all \textit{base words}. These two structures inform the global importance and weights of the words but are kept strictly separate, to guarantee a detachment between the user-defined concepts and the concrete topic modeling approaches. This, in turn, ensures model transferability and cross-corpus learning. 

\revision{We generically refer to all words (also n-grams) in the corpus, as well as words transitively contained in their embedding vectors,  as ``\textit{words}.'' 
\textbf{Base Words} are all words that are neither part of the higher levels of the}
\begin{wrapfigure}[7]{r}{0.43\columnwidth}
 \vspace{-20pt}
  \begin{center}
    \includegraphics[width=0.45\columnwidth, right]{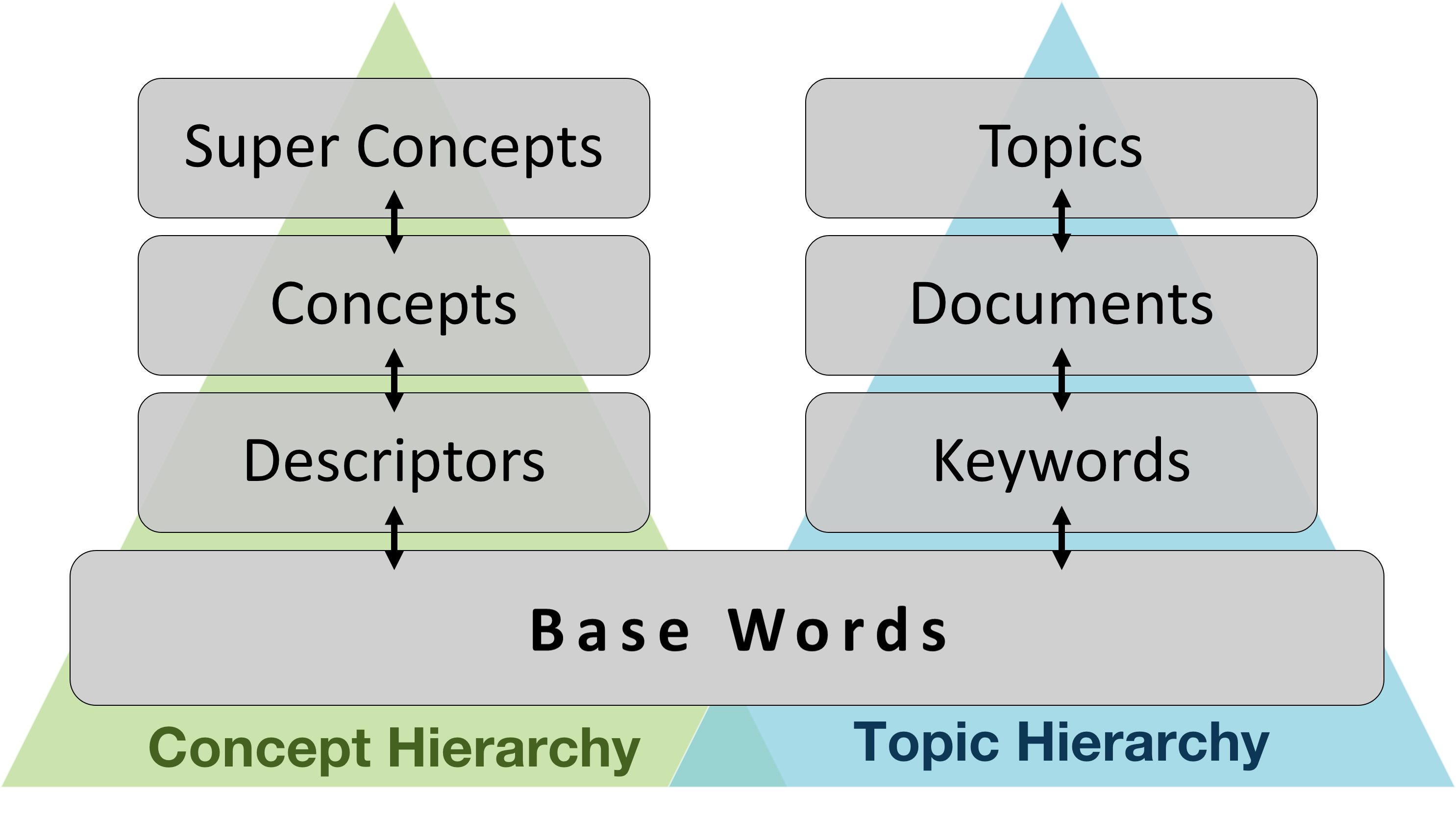}
  \end{center}
%    \vspace{-22pt}
\end{wrapfigure}
\revision{concept nor of the topic hierarchies. They can be promoted to become keyword and/or descriptors through user interaction. On the other hand, demoted keywords and/or descriptors traverse down the hierarchy to become base words. As suggested by their name, these form the basis of the two data hierarchies. } %, which are structured as follows. }

The \textbf{\concept{Concept Hierarchy}} \revision{ is user-driven and reflects the semantic relation between the words based on the domain knowledge externalization of users. \textbf{Descriptors} build the lowest level above the base word and are all the words that describe a concept (one level up) but that are not} 
\revision{concepts or super concepts themselves. Descriptors have a strict parent-child relation (1:n) to concepts. }
% \textbf{Descriptors} are all keywords that can be promoted to become concepts.
\textbf{Concepts} define the users’ semantics and are used as the main level of interaction. \revision{They are the link between the descriptors (their children) and the super concepts (their parents). Although a word can only be either a descriptor or a concept (exclusive relation), super concepts can \textit{include} concept words. The reason for this decision is that in some corpora there are multiple super concepts that only contain one concept each.} Hence, \textbf{Super Concepts} are automatically computed as a summary of the underlying region. \revision{Users can define the level of abstraction, i.e., the number of super concepts interactively. However, in contrast to concepts, the parent-child relationship between concepts and super concepts cannot be manually adjusted but is computed to give a faithful overview of the current state of the concept hierarchy at the time of viewing.}

\revision{In contrast, the} \textbf{\topic{Topic Hierarchy}}\revision{ is data-driven. It reflects the structure of the underlying corpus based on the selected topic modeling approach. 
\textbf{Keywords} are all words contained in all corpus documents, \textit{including} all descriptive document keywords. 
\textbf{Documents} are the given unit of analysis in a corpus and are each represented by their top $n$-keywords.
\textbf{Topics} are computed using a topic modeling algorithm and are each represented by their top $m$-keywords. Note, that the number of top keywords $n,m$ for document and topics, respectively, can be adjusted by the user. By default both parameters are set to 15 keywords. }

All words used in this approach are processed through a linguistic pipeline~\cite{El-Assady2017c,lingvisio}, that includes stemming, POS tagging, n-gram extraction, stop-word removal, and scoring. As described in our previous works~\cite{El-Assady2018ProgressiveFramework,el2018visual}, we treat each word as a weighted vector initialized using a user-selected scoring function~\cite{Manning1999}. %, e.g., tf-idf~\cite{sparck1972statistical}, log-likelihood ratio~\cite{Manning1999}, or G$^2$~\cite{rayson2000comparing}.
This section discusses the modeling of the semantic concept spaces, including the creation of the concept hierarchy.
The topic modeling hierarchy, on the other hand, is subject to the concrete algorithm used. \revision{Since our approach is independent of concrete topic modeling techniques, in this paper, we do not discuss the topic modeling process in detail. For more information on topic modeling and the concrete algorithm used throughout this paper, please refer to our previous work}~\cite{el2018visual}. \revision{Rather, in this paper, we, focus on the  \textit{model-agnostic} optimization of topic modeling through concept space refinement.
Section \ref{sec:ml} discusses this \textit{iterative} refinement process and the interplay between the concept and topic hierarchies, in more detail.}  Both hierarchies operate on the same underlying word vectors. Changes in the concept hierarchy, therefore, influence the scoring of words and, in turn, affect the topic modeling. 
\revision{This section discusses the four step process of modeling the underlying data structure of the \textit{semantic concept space}. This includes building the concept hierarchy, as well as deriving a spatialization of all objects in the concept and topic model views based on the relations of the underlying vector space model. This spatialization is used to initialize the two views, as described in Section~\ref{sec:va}. To facilitate the readability of this section, we use a simplified example of two generic agenda items from recent US presidential debates, namely, \textit{healthcare} and \textit{taxes}.}

\begin{figure*}
	\centering
	\includegraphics[width=\textwidth]{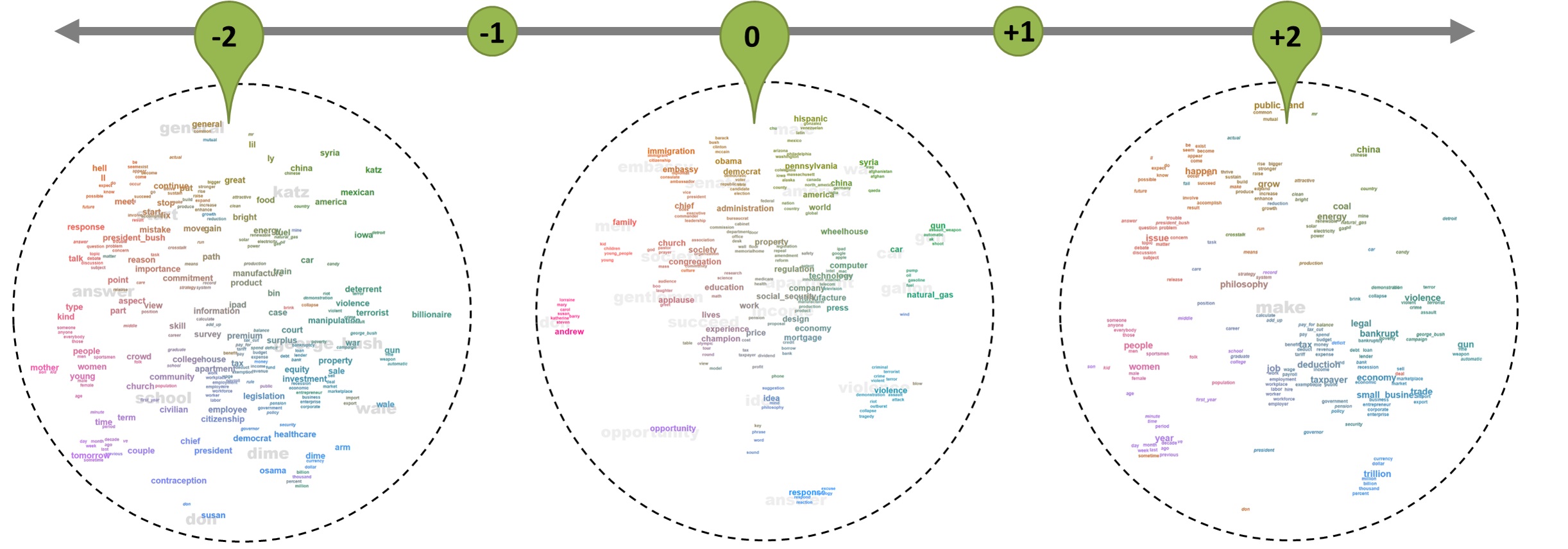}
	\caption{Semantic Abstraction Levels for Concepts.  By default, the entry point for the visualization (0) shows all major concepts. Users can opt to start at a lower abstraction level (-2), revealing more concepts, or choose a higher abstraction level (+2), resulting in fewer initially visible concepts. }
	\label{fig:AbstractionLevels}
	\vspace{-15pt}
\end{figure*}

\subsection{Interactive Concept Generation}
\label{sec:concept-generation}
The first step in this modeling pipeline is the generation of \textit{weighted concept vectors}. % as initial input. 
Assuming that the users' domain understanding can effectively guide the automatic computation in this initial step, we allow users to \textit{optionally} intervene and interactively edit suggested concept keywords which are used as priors for the further computation. This initial \revision{concept generation} is described in following four-step process: 

\textbf{(1) Seed Concept Extraction -- }
After pre-processing and annotating all the words in the document collection to be analyzed, 
% Before the concepts are calculated, 
we extract seed words. We rely on (a) \textit{Latent Dirichlet Allocation}~\cite{Blei2003LatentAllocation}, as well as a (b) \textit{Document Descriptor Extractor}~\cite{El-Assady2018ProgressiveFramework} to extract the most descriptive keywords in a corpus based on \revision{word frequencies}, tf-idf~\cite{sparck1972statistical}, log-likelihood ratio~\cite{Manning1999}, and G$^2$~\cite{rayson2000comparing} metrics. 
\revision{Note, that these two methods are only used as a heuristic for an initial fast separation of the overall corpus space. We do not apply LDA for topic modeling.}
The extracted seed words are considered the first concepts and are expanded in the next step to concept vectors.
\revision{In our example, this step might return two keywords like \textit{medical} and \textit{taxes}. }

\textbf{(2) Concept Vector Expansion -- }
In this step, the initial concept words are enriched with semantically similar words using the word embedding service \textit{ConceptNet}~\cite{speer2017conceptnet} to create \textit{concept vectors}. 
% All words contained in the enriched vectors that are not part of the %analyzed 
% corpus are discarded to focus the vector space and avoid skewness. 
\revision{Words that are not part of the corpus, but contained in an enriched vector are discarded to focus the vector space and avoid skewness.}
Note, that we extract word embedding vectors for all words in the %analyzed 
corpus but only vectors associated with \textit{concept} words are called \textit{concept vectors}. All words in a concept vector are regarded as \textit{descriptors} for their respective concept.
\revision{In our example, the concept vectors might contain the following descriptors: $\vec{medical}$\textit{:\textless system,health,relief,care\textgreater} and   $\vec{taxes}$\textit{:\textless deduction,money,cuts,relief\textgreater}.}

% \copied{The resulting words are immediately raised to concepts and enriched with similar words that are also found inside the corpus, using word embedding. }

\textbf{(3) Interactive Editing and Enrichment -- }
After the first two unsupervised steps, we involve the user in the concept generation. 
Similar to our proposed \textit{topic backbone}~\cite{el2018visual}, users have the option to adjust the seed concepts and their vectors as they see fit. They can as well introduce new concepts or remove descriptors to adapt the generated concepts to their understanding. However, as we cannot always assume that users have existing knowledge about the corpus before exploring the visualization, this processing  step is optional. If skipped, the concept vectors from the previous step will remain unchanged. \revision{A user might, for instance, choose to add the descriptor \textit{healthcare} to $\vec{medical}$. }

\textbf{(4) Scoring and Ranking -- }
After the generation of the \textit{concept vectors}, in this step, we use the scoring functions from the \textit{Document Descriptor Extractor}~(1b) to rank the descriptors of each concept. The ranking and scores of each concept is used for weighting them later on.
\revision{For instance, the words \textit{system} and \textit{relief} in our example concepts could be ranked low as these words are, in the one case, too generic and, in the other, too undescriptive (i.e., occurring in both concepts).  }

\subsection{Concept Neighborhood Computation}
\label{sec:concept-neighborhood}
Based on the weighted concept vectors, this step computes semantic concept neighborhoods to determine the spatialization of all the words in the analyzed corpus. The output of this step is, therefore, a set of \textit{2D-coordinates \{x,y\} for each word}, anchored by concept neighborhoods.  
%
% \revision{Anchors are the initial seed points for concepts and their descriptors}
We rely on \textit{t-distributed Stochastic Neighbor Embedding} (t-SNE)~\cite{hinton2003stochastic} for the computation of the concept neighborhoods based on the word embedding vectors. To guarantee a more stable projection result we use the \textit{concept vectors} as anchors throughout this work. Furthermore, we configure the t-SNE calculation with the following parameters; a perplexity of $5$, a theta of $0.5$ and $5000$ learning iterations. These were determined based on trials with different corpora using a projection inspection approach~\cite{SKJE18}. As the perplexity parameter describes the expected minimum number of neighbors each point should have, to ensure a convergence with few errors (i.e., separable while preserving object distances), it is essential to maintain a partly overlapping set of descriptors in the enriched concept vectors. %have a common only use very few concepts but have them enriched with similar concept descriptors
In the following, we describe the three-step process for computing semantic concept neighborhoods.

\textbf{(1) Corpus and Topic Keyword Insertion -- }
In order to consider all relevant words in the projection, in this first step, we \revision{combine} all corpus and topic keywords (each represented by their word embedding vector) with all extracted concepts (represented by their respective concept vectors). We use all word embedding vectors in the second step to determine the initial positioning of the concept vectors. These positions, in turn, are used as anchors in the third step.  

To ensure that the concept space is representative of the analyzed corpus, in this step we additionally assign the top twenty keywords from each document to their closest concept vector as descriptors. \revision{In our example, we might add keywords like \textit{company} or \textit{spending} to $\vec{taxes}$, as well as \textit{affordable} to $\vec{medical}$.}
Note that this technique is independent of the concrete topic modeling approach, as long as each topic is represented by a keyword vector and each document is assigned to a topic. In this paper, we use the \textit{Incremental Hierarchical Topic Model} (IHTM)~\cite{el2018visual} throughout, as it is deterministic and provides the required topic-document-keyword hierarchy.

\textbf{(2) Initial Concept-Anchor Setting -- } To meaningfully initialize the t-SNE projection, in this step we compute \textit{\{x,y\}-coordinates} for all extracted concepts and set these as anchors for the projection in the next step. We determine these coordinates based on a run of t-SNE on the complete set of word vectors in the corpus. This first run uses random initial positions for the words as it is only employed to determine a meaningful spatialization for the semantic concepts. Therefore, other than the \textit{\{x,y\}-coordinates} for the concepts, the word positions of this run are discarded and recalculated in the next step.  

\textbf{(3) t-SNE Reduction  -- }
To retain a stable projection after t-SNE convergence, in this step, we use the previously determined concept positions as anchors. We then run t-SNE a second time to determine the \textit{\{x,y\}-coordinates} for all word vectors in the space. %These include all corpus words regardless of their association to specific concepts or topics. 
In later steps, when users edit and change the concept hierarchy, we re-run this step on-demand to update the concept space. 
\revision{In our example, each of the two concept vectors, as well as their associated descriptors have a determined position as \textit{\{x,y\}-coordinates} in the 2D space.}

\subsection{Concept Hierarchy Building}
Based on the neighborhoods determined by the word embedding projection, in this step, we build the \textit{concept hierarchy relations}, getting rid of descriptor overlap by assigning each descriptor to only one concept. 
This is achieved based on the following four-step process:

\textbf{(1) Parameter and Constraint Setting -- }
The abstraction level of the concept space has a considerable impact on the visual analysis and refinement process. We therefore present users with a choice of different entry points in the visual analytics interface. We provide non-overlapping level-of-abstraction sliders to adjust the  \textit{semantic abstraction levels} for concepts and super concepts.  For example, \autoref{fig:AbstractionLevels} shows three out of five abstraction levels for concepts. 

The parameters chosen to determine these abstraction levels are two-fold: The minimum semantic \revision{cosine} similarity threshold $\varepsilon_{similarity}$, and the minimum number of descriptors or concepts in a neighborhood 
\begin{wrapfigure}[4]{r}{0.21\columnwidth}
 \vspace{-24pt}
  \begin{center}
    \includegraphics[width=0.23\columnwidth, right]{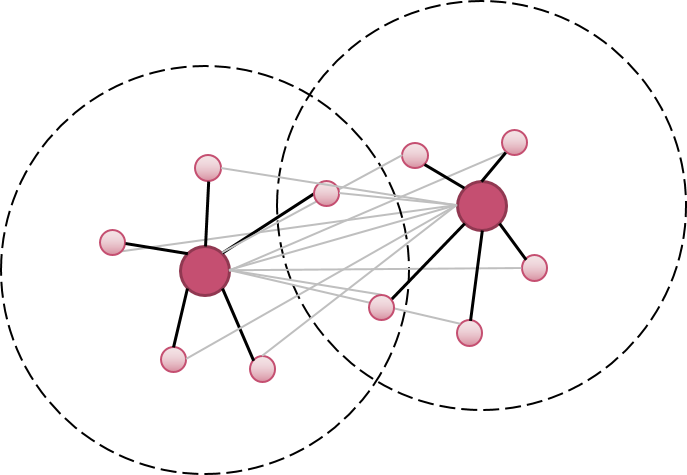}
  \end{center}
%    \vspace{-22pt}
\end{wrapfigure}
$\varepsilon_{neighborhood}$.
By default the similarity threshold is set to $\varepsilon_{similarity} = 0.4$, and the neighborhood parameter is set to $\varepsilon_{neighborhood} = 6$ for concepts and to $1.5  \times  \varepsilon_{neighborhood}$ for super concepts. Changing the abstraction slider adjusts  $\varepsilon_{neighborhood}$, directly resulting in a higher or lower level of abstraction. 
Based on these parameters we perform a hierarchical, density-based clustering to obtain the concept hierarchy, as described in the next steps.  

\textbf{(2) Semantic Similarity Update -- }
Beside the word positioning, to perform the hierarchical clustering, we use the above mentioned semantic similarity. 

\revision{We keep words inside the same cluster (i.e., concept) if they are similar with respect to their \textit{cosine similarity}. 
We check this at two points during clustering. 
First, when deciding which words could initially form a cluster using the $\varepsilon_{similarity}$ threshold. 
Secondly, when clusters overlap, only clusters which have a high word-embedding coherence are merged, Otherwise, all overlapped members are redistributed to their most similar cluster. 
The word-embedding coherence defines the threshold for the minimal acceptable inter-cluster coherence on the current abstraction level, and is dependant on the $\varepsilon_{similarity}$ and the current concept abstraction level.
Hence, updating the semantic similarity based on the $\varepsilon_{similarity}$ threshold is essential to ensure a coherent semantic concept hierarchy.}
\revision{In our example, the word \textit{system} and the concept \textit{medical} might not meet the $\varepsilon_{similarity}$ threshold. }

\begin{wrapfigure}[6]{r}{0.23\columnwidth}
 \vspace{-28pt}
  \begin{center}
    \includegraphics[width=0.25\columnwidth, right]{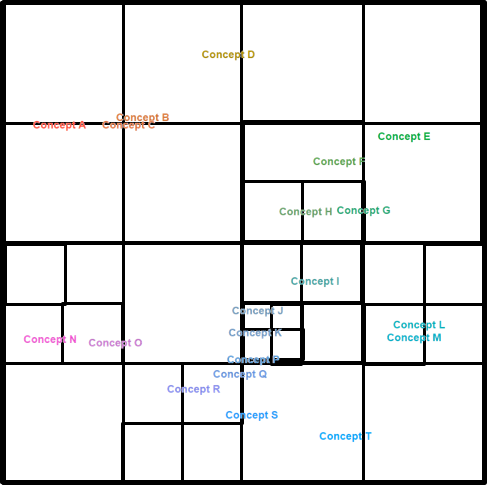}
  \end{center}
%    \vspace{-22pt}
\end{wrapfigure}
\textbf{(3) Quadtree Mesh Generation -- }
The second 
criterion used in the clustering is neighborhood preservation. Based on the \textit{\{x,y\}-coordinates} previously obtained for each word, we generate a \textit{quadtree}~\cite{Finkel1974} mesh, such that every word is positioned in its own quadrant. The quadtree recursively partitions the 2D space into squares, where each non-empty square is further divided into four equal-sized squares. Hence, each point (i.e., word) has its own leaf node. Coincident points are stored as a linked list. The quadtree is used in later steps \revision{as an index} for  collision detection and neighborhood search. 

\textbf{(4) Hierarchical Density-Based Clustering -- }
\label{sec:concept-clustering}
Based on the $\varepsilon_{similarity}$ and $\varepsilon_{neighborhood}$ thresholds, we calculate the concept hierarchy in two separate clustering iterations, one for concepts and another 
\begin{wrapfigure}[7]{r}{0.23\columnwidth}
 \vspace{-20pt}
  \begin{center}
    \includegraphics[width=0.25\columnwidth, right]{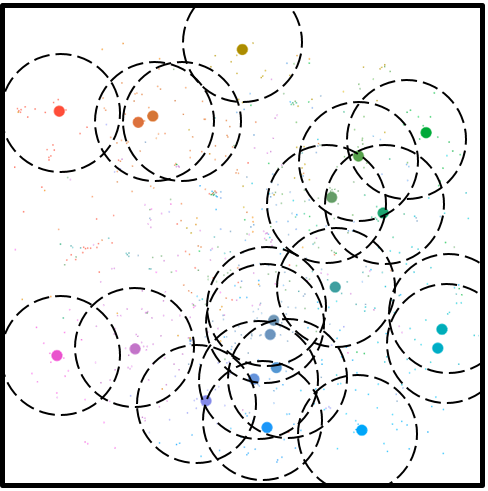}
  \end{center}
%    \vspace{-22pt}
\end{wrapfigure}
for super concepts. We perform an \textit{agglomerative, density-based clustering}~\cite{Zhao2005} that assigns each word in the space to a concept, and, in turn, each concept to a super concept. Based on the quadtree, we extract the $\varepsilon_{neighborhood}$ nearest neighbors of each concept word and form initial concept clusters. If a concept does not have enough neighbors ($\varepsilon_{neighborhood}$) or these neighbors do not satisfy $\varepsilon_{similarity}$, we do not create a cluster.  Once two clusters overlap, their overlapping children are either split up or the two clusters are merged (depending on their pairwise \revision{cosine} similarity).  After the initial clustering is formed, all  descriptors not belonging to a cluster are assigned to their most similar concept. The process is repeated to group concepts into super-concepts using their respective parameters. %(Chris: perhaps no need to say what isn't done) Note, that words in the space that are neither descriptors nor have been considered by the clustering are not assigned to a particular concept -- only if interactively inserted by the user do they get a concept assignment. 
\revision{After clustering, our example concepts %could
become: $\vec{taxes}$\textit{:\textless cuts, deductions, spending, company\textgreater} and $\vec{medical}$\textit{:\textless healthcare,health,care,affordable\textgreater}. Note, that for $\vec{medical}$ the word \textit{health} is added, as it is among its nearest neighbors.   }

\subsection{Layered Canvas Mapping}
Based on the spatialization of all words and the generated concept hierarchy, the last step in modeling the semantic concept space is the \textit{layered mapping  of all elements on a canvas}. 
In the following, we describe the five-step layout process. %, which is repeated every time the concept space is recomputed. 

% Output: canvas with mapping of elements

\textbf{(1) Transformation and Rescaling -- }
To maximize available screen space, all data points are transformed and rescaled to the boundary of the rectangular viewport.  %We apply a uniform, rectangular distribution  
The canvas on which the points are mapped can then be interactively panned and zoomed by  users during analysis.

\begin{figure*}
\centering     
\subfigure[\textbf{\concept{Concept View}} showing an overlay of all visualization layers.]{\label{fig:ConceptView}\includegraphics[width=0.495\textwidth,cfbox=light-gray 1pt 0pt]{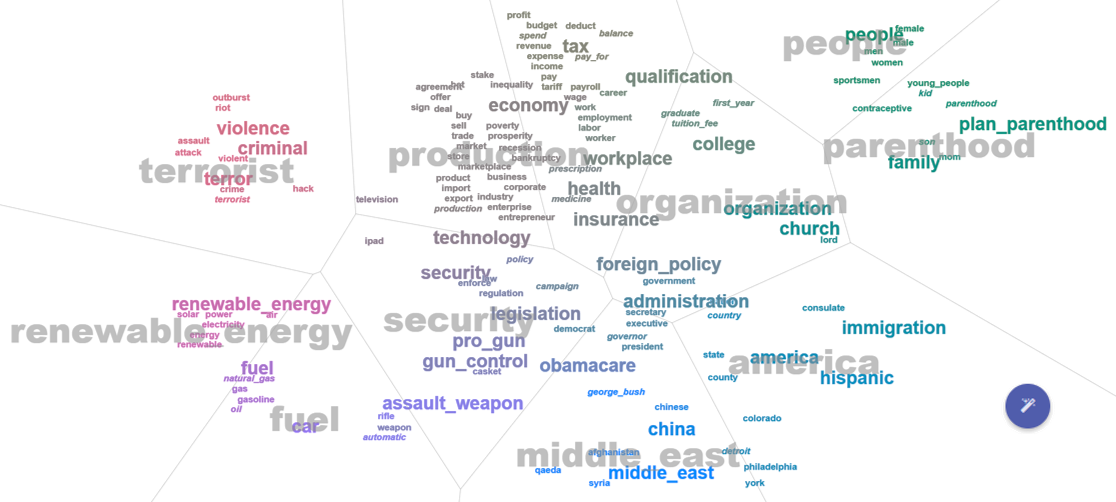}}
\subfigure[\textbf{\topic{Topic View}} depicting two selected \textit{topic glyphs}.]{\label{fig:TopicView}\includegraphics[width=0.49\textwidth,cfbox=light-gray 1pt 0pt]{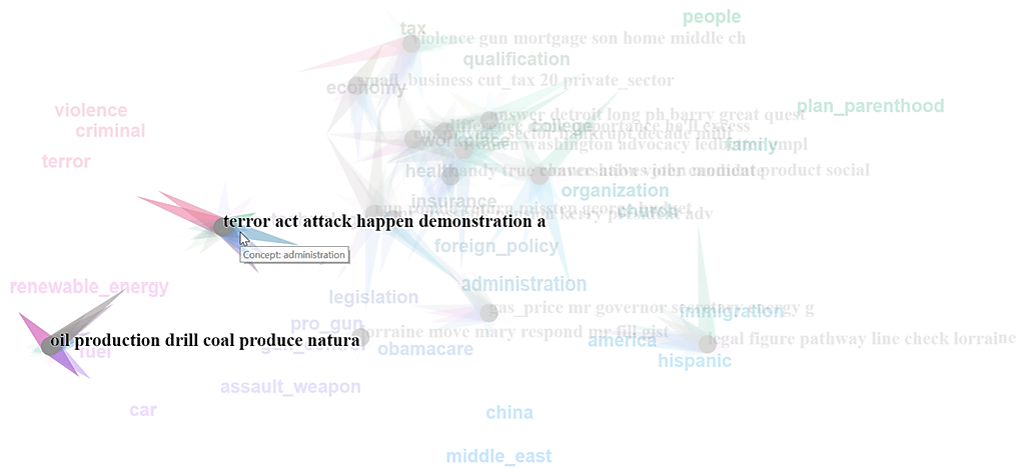}}
\caption{Duality of Concept and Topic Views. Selected layers from each view `\textit{shine through}' the other view to give context. In this example, the left side of the (a)~\concept{\textbf{concept view}} represents a region on \textit{renewable energy} (bottom) and \textit{terrorism} (top), while the corresponding (b)~\topic{\textbf{topic view}}  places the topic on \textit{oil production} to \textit{renewable energy} and the topic on a \textit{terror attack in Libya} between the two concepts as it is related to both.} %more than one aspect.  }
	\label{fig:Overview}
	\vspace{-15pt}
\end{figure*}

\textbf{(2) Concept-Anchored Projection -- }
Based on the rescaled canvas, as well as the adjusted concept hierarchy, we recompute a concept-anchored t-SNE reduction and project the result onto the canvas. In addition to the words being projected as points based on their respective \textit{\{x,y\}-coordinates}, we include a \textit{bounding box} for every word based on its length and size (i.e., concepts are shown larger than descriptors, etc.)  The quadtree index is also updated during this step.

\textbf{(3) Overlap Reduction -- }
Since some of the word vectors might be projected onto close coordinates on the canvas, in this step we reduce the potential overlap. Iterating over the quadtree index, we detect areas of potential occlusion based on the object position, as well as its bounding box. The overlapping objects are moved away from each other until the process has reached a local minimum.

\textbf{(4) Color Mapping -- }
The layout process results in 
\begin{wrapfigure}[7]{r}{0.23\columnwidth}
 \vspace{-20pt}
  \begin{center}
    \includegraphics[width=0.25\columnwidth, right]{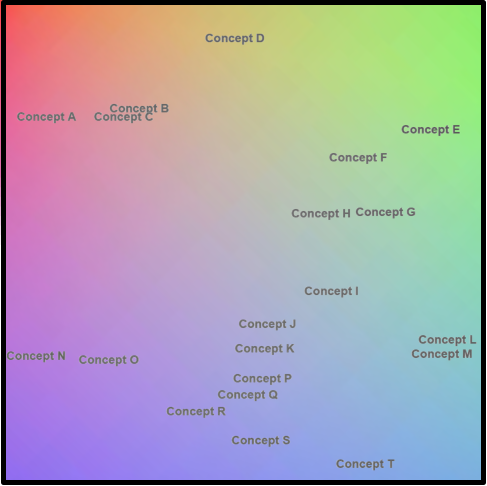}
  \end{center}
%    \vspace{-22pt}
\end{wrapfigure}
\textit{\{x,y\}-coordinates} which reflect semantically similar neighborhoods.  
In addition to this spatial  encoding, we use the \textit{LAB Color Space}~\cite{wyszecki1968color} to assign each concept a color based on its respective 2D-position in that space. Descriptors are assigned the colors of their parent concept. This double encoding of similarity reveals descriptors that are projected in a neighborhood of a different color, indicating that their underlying word embedding is in conflict with the concept hierarchy. 

\textbf{(5) Voronoi Tessellation -- } To structure the concept space, we rely 
\begin{wrapfigure}[7]{r}{0.23\columnwidth}
 \vspace{-20pt}
  \begin{center}
    \includegraphics[width=0.25\columnwidth, right]{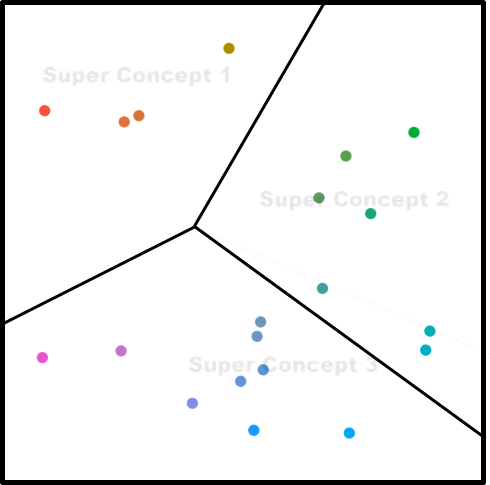}
  \end{center}
%    \vspace{-22pt}
\end{wrapfigure}
on the positioning of the extracted concept hierarchy. To enhance the visual association of the words in the space to super concepts (creating a high-level overview), we partition the space based on the extracted super concepts. We employ a \textit{Sweepline Voronoi Algorithm}~\cite{fortune1987sweepline} to determine super concept boundaries which can be visualized on the concept space canvas on demand.

% https://github.com/d3/d3-voronoi

%********************************************

\section{Visual Analytics Workspace}
\label{sec:va}
The generated data structures and spatialization, described in \autoref{sec:modeling}, builds the foundation of the visual analytics workspace. 
\begin{wrapfigure}[8]{r}{0.21\columnwidth}
 \vspace{-24pt}
  \begin{center}
    \includegraphics[width=0.23\columnwidth, right]{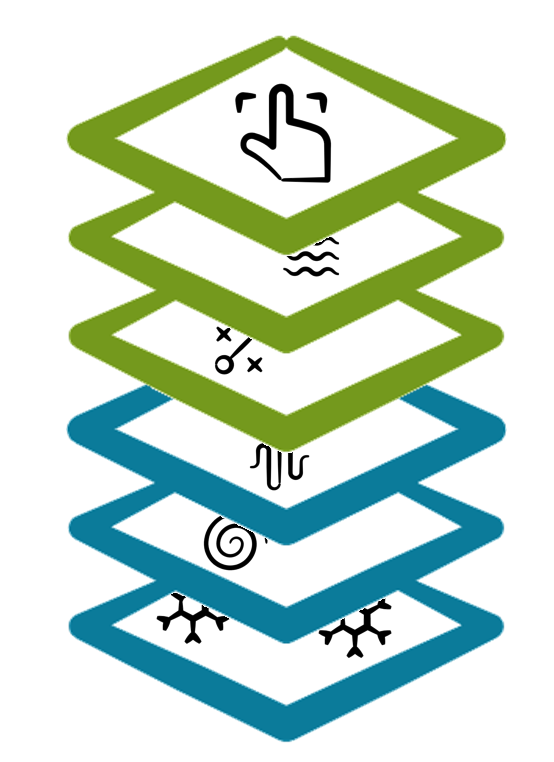}
  \end{center}
%    \vspace{-22pt}
\end{wrapfigure}
The \textit{Semantic Concept Space} is designed as a layered, interactive canvas that consists of two stacks of layers; (1)~the \textbf{\concept{Concept View}} and (2)~the \textbf{\topic{Topic View}}.

\textbf{Design \revision{Rationale} -- } Building on the basis of the word positioning, we designed the two views to be separate, super-positioned canvases. Users can interact with one view at a time, while the other is toggled inactive. To facilitate comparison between views, the inactive view is shown with a low opacity in the background of the active one, making its elements \textit{shine through} the canvas.  
Each view is composed of three layers, representing its hierarchy levels above the base words. The layers of the \concept{Concept View} are initialized with the extracted concept hierarchy, and the \topic{Topic View} layers with the topic modeling result. 
A concept refinement process enables users to  \revision{directly} adapt the concept hierarchy to their semantic knowledge. They can promote words up the hierarchy (base word $\rightarrow$ descriptor $\rightarrow$ concept) or demote them. Only the super concept layer is not interactively adjustable, as it is supposed to reflect a high-level view of the complete space. At any point \revision{of this iterative process} can users trigger a \textbf{recomputation of the t-SNE projection} to adjust the word spatialization to the new concept hierarchy.
On the other hand, the topic modeling view can \textit{not} be adjusted \revision{directly} but is used for inspecting and analyzing the topic modeling result. Only through \textbf{recomputing the topic modeling algorithm} (on-demand) do the layers of the topic modeling change to adapt to the concept refinements. This \textit{duality} of views enables users to \textit{teach} the machine learning model their domain knowledge, as well as the machine learning model to respond through \textit{learning} the new semantics.

\textbf{Visual Encoding -- } 
Our visual workspace is designed to support (1)~finding different elements on the canvas, %(1)~the localization of the different elements on the canvas, 
as well as (2)~the spatial association of words. As a second level task users are expected to (3)~decode the type of word object at hand. To design an appropriate visual encoding we consulted a study of the design space of keyword summaries~\cite{Felix2018TakingSummaries} that indicates that there is a trade-off in the effectiveness of typography versus marks with respect to our tasks; search speed~(1\&2) and value judgement~(3). According to their findings font size attracts the attention of users and performs better in search tasks. In addition, according to Alexander et al.~\cite{ale2017a}, the perceptual bias for estimating font sizes is negligible. We, hence, represent each word object by default with a \textbf{label} %\inlinegraphics{encoding4.jpg} 
and enable users to toggle on a
\textbf{circle} %\inlinegraphics{encoding3.jpg} 
as an additional mark.  Both the circle and the label sizes encode the object level in the data hierarchy and, thus, are doubled going up the hierarchy.  Furthermore, for the topic view, we designed a \textbf{topic glyph}~\inlinegraphics{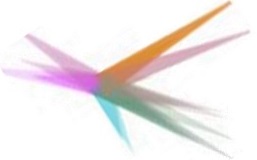} that represents the topic or document association with different concept regions. This glyph % \inlinegraphics{encoding2.jpg} 
can be used as another alternative representation for the object marks on the canvas layers.

\subsection{Concept View}
\label{sec:concept-view}
The concept view, Figure \autoref{fig:ConceptView}, is the entry point to the visual analytics workspace.  As shown in \autoref{fig:AbstractionLevels}, users can vary the
\begin{wrapfigure}[9]{r}{0.33\columnwidth}
 \vspace{-17pt}
  \begin{center}
    \includegraphics[width=0.35\columnwidth, right]{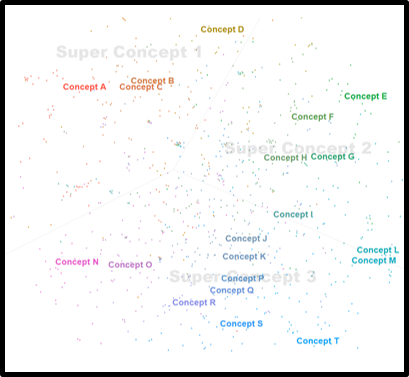}
  \end{center}
%    \vspace{-22pt}
\end{wrapfigure}
semantic
abstraction level of concept and super-concepts. %(c: sliders already described) view using two sliders; one for concepts %(C: number of steps doesn't matter) (with five steps) and the other for super concepts (with seven steps). 
Defining the entry point of their analysis is equivalent to choosing a \textit{refinement strategy}. Some users prefer to start at a detailed level and refine the concept space by deleting non-descriptive words (\textit{bottom-up refinement}), while others prefer to add descriptors (\textit{top-down refinement}) to an initially abstract view. After configuring the initial concept view, users can start exploring and interacting with the concept layers, as described in the remainder of this section.

\textbf{Descriptor Layer -- } There are three types of descriptors in the concept hierarchy; (1) descriptors from the concept generation step (\autoref{sec:concept-generation}), called \textit{concept descriptors}; (2) descriptors from the neighborhood computation step (\autoref{sec:concept-neighborhood}), called \textit{topic descriptors}; and (3) \textit{user-defined descriptors}. Each of these descriptors is directly assigned to a concept. Users can toggle the visibility of each of the descriptor groups. When visible, descriptors can be represented by a colored dot and/or a small label.% people know what a word is..., e.g.,   \inlinegraphics{descriptor3} or \inlinegraphics{descriptor2}. 
The color of a descriptor is based on the position of its parent concept, while its position is based on its weighted word embedding vector. This enhances the detection of outliers, i.e., as colors directly reflect the user-refined concept hierarchy. %, they make the \textit{folds} in the \revise{linguistic fabric} (acquired through domain knowledge externalization) visible and identifiable. 
% \inlinegraphics{descriptor1}

A single descriptor can be selected and deselected through a toggle-click. Selections can be (1)~\textit{deleted} from concept view, i.e., demoted to become a word; (2)~\textit{promoted} to become a concept; or (3)~\textit{(re-)assigned} to an existing concept. A group of descriptors can also be used to create a concept, promoting a selected one to a concept and all others as its new descriptors. In addition, users can \textit{add a word} to the descriptor layer, effectively promoting it to become a descriptor.

\textbf{Concept Layer -- } This is the central layer for the refinement of the concept hierarchy. Concepts are represented by a colored dot and/or a medium-sized label.%, e.g., \inlinegraphics{concept1}. 
 In contrast to descriptors, the color and position of concepts are synchronized in order to anchor the concept space. %(C: normal behavior of a vis) When selected, a concept and all its descriptors are highlighted, while the remaining elements in the space get faded out. 
A single concept can be (1)~\textit{demoted} to become a descriptor, redistributing its descriptors; (2)~\textit{deleted} together with all its associated descriptors (becoming base words); or (3)~\textit{swapped} by one of its descriptors.  A selected group of concepts can be (4)~\textit{merged} to form one joint concept.

\textbf{Super Concept Layer -- } The highest level abstraction in the concept hierarchy is formed by super concepts. These are represented by large, faded-gray labels positioned in the background.%, e.g., \inlinegraphics{superconcept1}. 
Super concepts are generated automatically and are non-interactive. However, users can vary the super concept abstraction level.  In addition to the labels, as described in \autoref{sec:concept-clustering}, super concepts structure the space into subdivisions. These are represented by a Voronoi tessellation.

% \newpage
\subsection{Topic View} 
\label{sec:topic-view}
To build this view, Figure \autoref{fig:TopicView}, %as a counterpart to the concept view, 
we rely on a spatialization of the keyword vectors of documents and 
topics. These vectors consist of a 
\begin{wrapfigure}[10]{r}{0.33\columnwidth}
 \vspace{-18pt}
  \begin{center}
    \includegraphics[width=0.35\columnwidth, right]{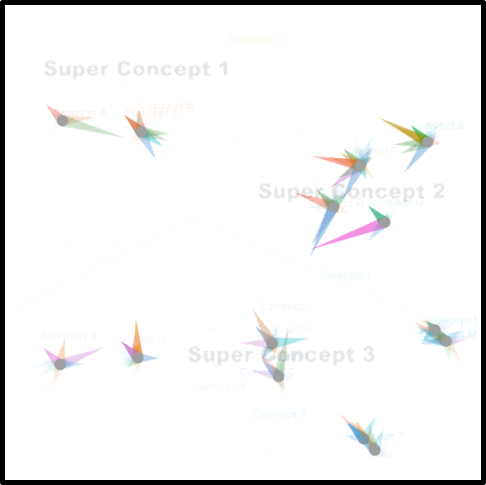}
  \end{center}
%    \vspace{-22pt}
\end{wrapfigure}
weighted set of the most descriptive keywords extracted by the underlying topic model~\cite{el2018visual}. These weights represent the importance of a keyword to their respective document or topic. In addition, every keyword has its own word embedding vector and weights, corresponding to its global importance in the corpus. The former weight is the \textit{learned} weight by the topic modeling algorithm, while the latter is influenced by the concept hierarchy manipulation to \textit{teach} the model. This section describes %the three topic view layers, % 
all %three 
layers of the topic view, 
as well as, the \textit{topic glyph} design.

\textbf{Keyword Layer -- } This layer corresponds to the descriptor layer of the concept view. Keywords are all descriptive words extracted by the topic modeling. %algorithm. 
Some of them are at the same time descriptors in the concept hierarchy. They are represented by a black label and/or a circle. %, e.g., \inlinegraphics{keyword} or \inlinegraphics{keyword1}.  
 Keywords are assigned to documents.  However, in contrast to the strict descriptor--concept assignment,  in topic hierarchy, more than one document can share the same keyword. Other than showing or hiding them from the canvas, keywords are not interactively adjustable.  
% \inlinegraphics{keyword2}

\textbf{Document Layer -- } The main unit of analysis in a corpus are the documents.  These are represented by their most descriptive keywords.%, e.g.,  \inlinegraphics{document}. 
In addition to circles and labels, documents can also be depicted using the \textit{topic glyph}. This indicates all related concept regions of a document, as described at the end of this section. Selecting a document reveals all its corresponding keywords. If one of its keywords is also a descriptor it gets highlighted in color, otherwise, document keywords are shown in gray. Moreover, hovering over a document object shows the underlying text for \textit{close-reading}. 

\textbf{Topic Layer -- } The top layer in this view is the topic layer. Similar to documents, topics are depicted by their top keywords. %, e.g., \inlinegraphics{topic}. 
They are also represented by labels, circles, as well as topic glyphs. Selecting a topic shows all the documents assigned to it.

\textbf{Topic Glyphs -- } To facilitate the association of topics and documents 
to concepts, the topic glyphs \revision{relate both} %set both in relation 
though spikes that point
\begin{wrapfigure}[11]{l}{0.35\columnwidth}
 \vspace{-25pt}
  \begin{center}
    \includegraphics[width=0.35\columnwidth, right]{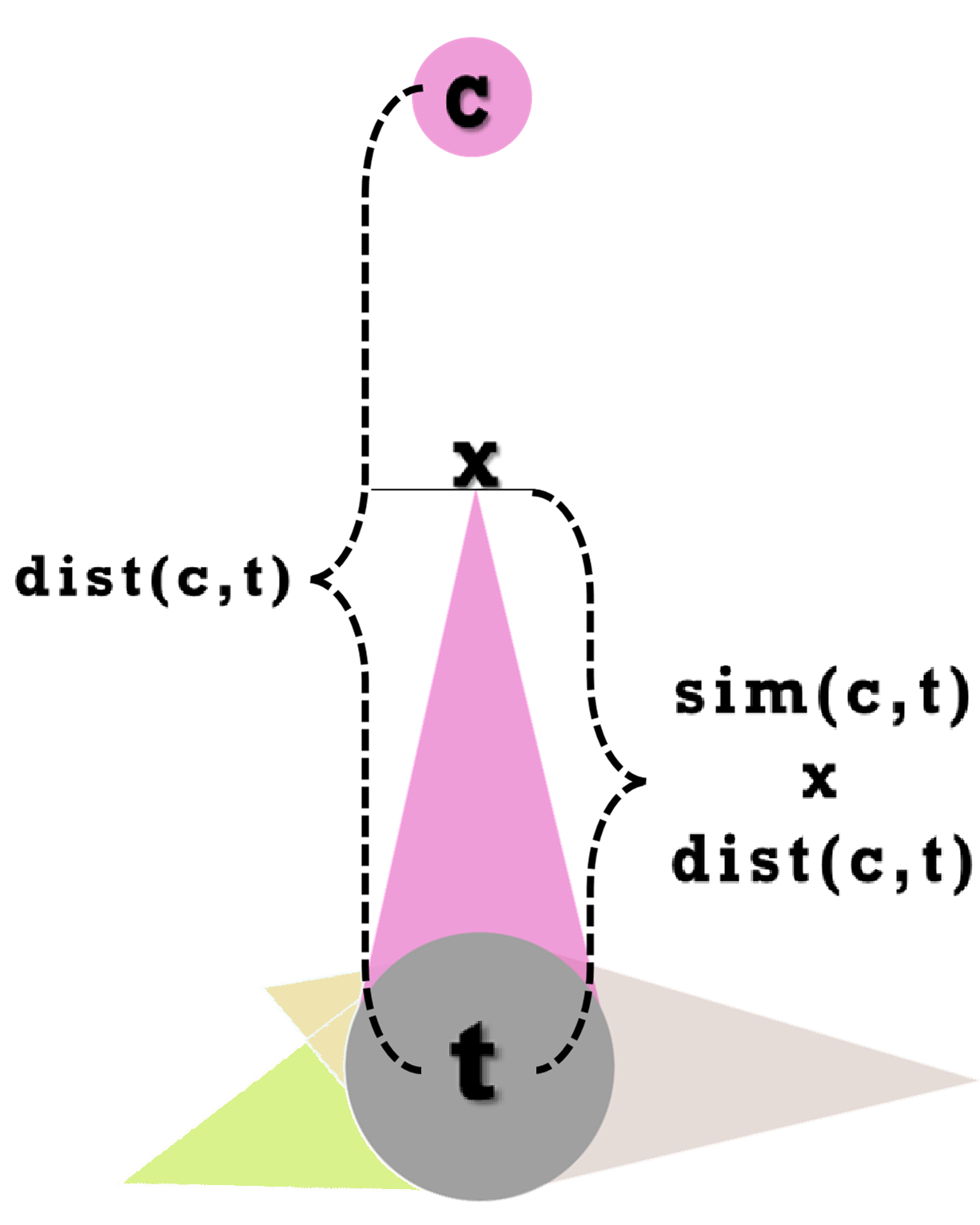}
  \end{center}
%    \vspace{-22pt}
\end{wrapfigure} 
to \revision{their} most related concepts in the embedding space. For each concept in the concept hierarchy, we include one spike such that the \textit{percentage of how similar} a given topic $t$ and the concept $c$ is proportional to the \textit{length} of the spike. We calculate the Euclidean distance between the two objects as $dist(c,t)$, and the normalized cosine similarity of their respective word embedding vectors as $sim(c,t)$. The distance marking the end point of the spike $x$ is thus the normalized product of the two factors: $dist(t,x) = sim(c,t) \times dist(c,t)$. %, as shown in~\autoref{fig:glyph}. 
In addition to being the scaling factor for each spike's length, the cosine  

\noindent similarity  $sim(c,t)$ is also mapped to  
the opacity of the spike, making 
\begin{wrapfigure}[11]{r}{0.14\columnwidth}
 \vspace{-24pt}
  \begin{center}
    \includegraphics[width=0.18\columnwidth, right]{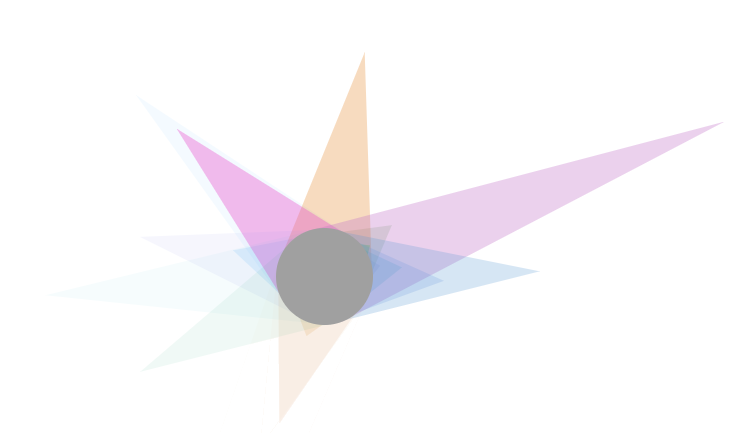}
     \includegraphics[width=0.18\columnwidth, right]{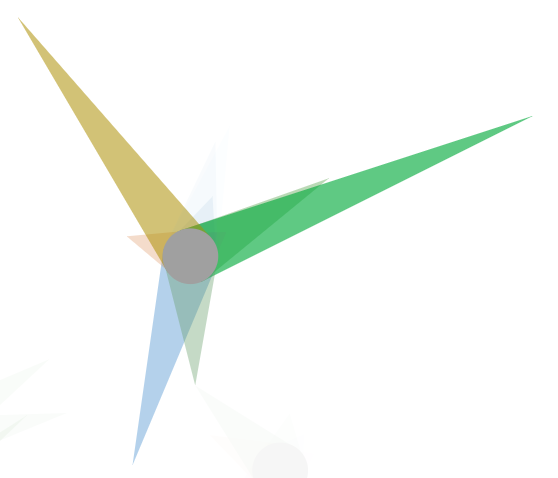}
    \includegraphics[width=0.18\columnwidth, right]{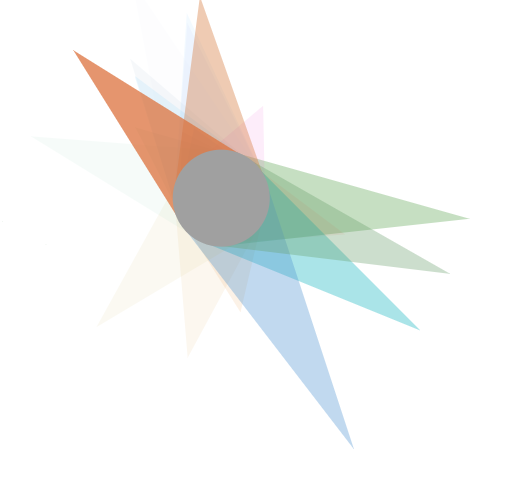}
  \end{center}
%    \vspace{-22pt}
\end{wrapfigure} 
the ones pointing to \textit{similar but distant} concepts more \textit{visually prominent}. To facilitate finding the locations of the concepts associated with a glyph spike, we map the color of each spike with the color of the corresponding concept. We further orient the spikes to point to their associated concepts. 
Hence, using this representation, we can reveal topics and documents that are mixtures of different concepts. Some of which, are intentionally bringing together aspects of two concepts, e.g., the topic on the \textit{attack on Libya} in  Figure~\autoref{fig:TopicView}, bringing together the concepts \textit{terrorism} and \textit{oil production}.  In other cases, topics are only associated with one concept and are correctly placed atop that concept, showing almost no visible spikes. Overall, these spikes can be seen as the directions in which topics or documents are \textit{pulled}, based on the semantics of the concept space.

\begin{figure*}
\centering     
\subfigure[\textbf{Direct manipulation} after lasso-selection of descriptors.]{\label{fig:DirectManipulation}  \includegraphics[width=.218\textwidth,cfbox=light-gray 1pt 0pt]{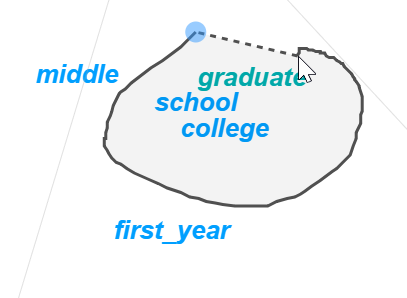}
 \includegraphics[width=.23\textwidth,cfbox=light-gray 1pt 0pt]{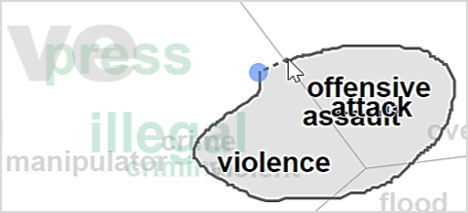} }
\subfigure[\textbf{Guided relevant feedback} suggesting to switch descriptor and concept.]{\label{fig:GuidedInteraction}
    \includegraphics[width=0.25\textwidth,cfbox=light-gray 1pt 0pt]{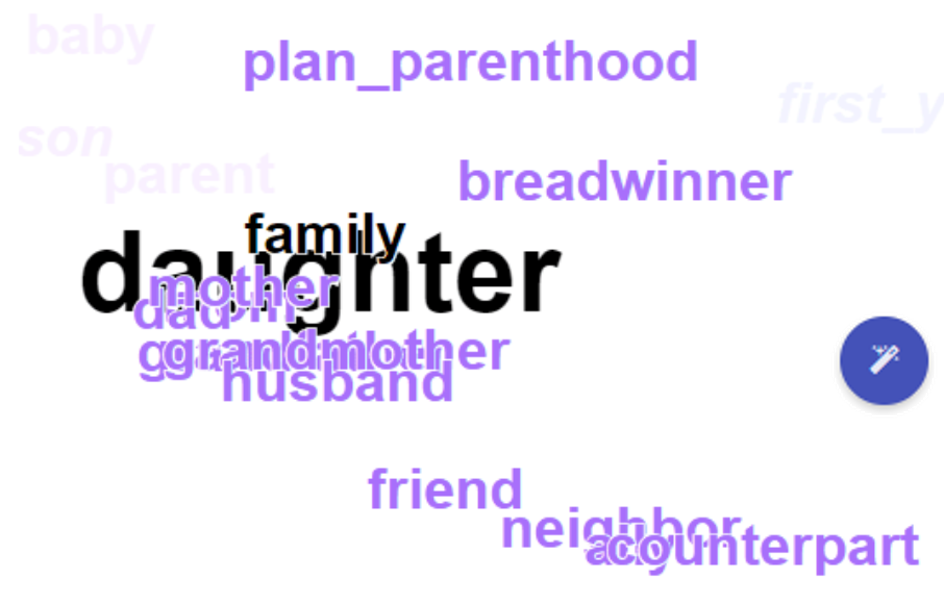}
    \includegraphics[width=0.25\textwidth,cfbox=light-gray 1pt 0pt]{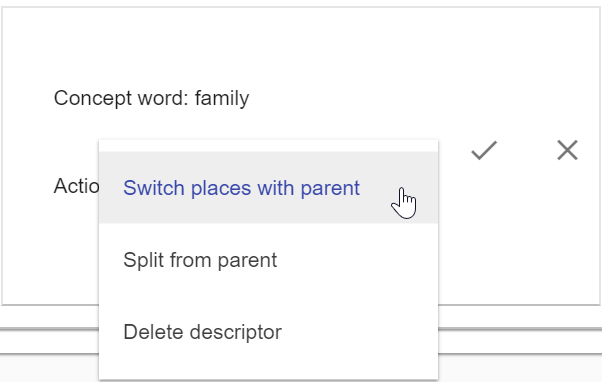}}
    \vspace{-5pt}
\caption{Two options for Concept Refinement. The direct manipulation enables exploratory refinement, while the guided relevance feedback is designed for targeted refinement. Both options can be used anytime throughout the visual analytics process to adjust the concept hierarchy.}
	\label{fig:Refinement}
	\vspace{-15pt}
\end{figure*}

\subsection{Concept Space Interactions}

Users are supported in their exploration and analysis through a number of \textit{instruments}.  The most noteworthy interactions are:

\textbf{Navigation through Word Search -- } We provide a search query interface and base words that are not found in the current hierarchy can be added as new descriptors.  

\begin{wrapfigure}[6]{r}{0.43\columnwidth}
 \vspace{-9pt}
  \begin{center}
    \includegraphics[width=0.45\columnwidth, right]{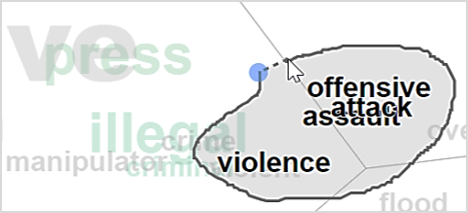} 
  \end{center}
\end{wrapfigure}
\indent\textbf{Lasso Selection -- } To facilitate the selection of multiple objects 
in the space with one interaction, we implemented a lasso selection. Individual items can be added to, or removed from existing selections.

\textbf{X-Ray Lens -- } 
Users might want to inspect a neighborhood
\begin{wrapfigure}[7]{l}{0.33\columnwidth}
 \vspace{-25pt}
  \begin{center}
    \includegraphics[width=0.37\columnwidth, left]{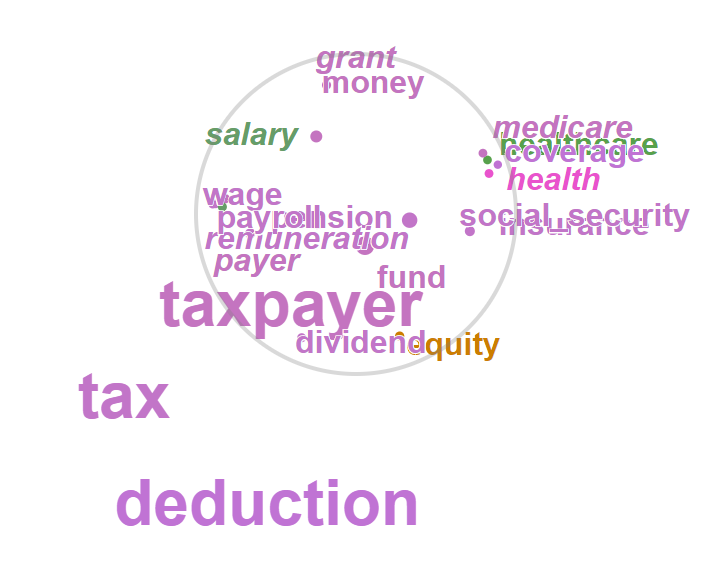}
  \end{center}
%    \vspace{-22pt}
\end{wrapfigure}
 in 
more detail, for example, to look for specific words, or to understand why an area might be empty in a specific layer. We, therefore, enable them to peek through \revision{all the layers at once} %the \revise{linguistic fabric} 
using a distortion lens. When activated in one of the views, the lens can be used to reveal all objects in a particular position throughout the hierarchy. The lens only operates on the active view. 

\textbf{Guided Tours -- } Lastly, users can toggle the `\textit{magic wand}' icon~\inlinegraphics{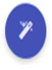} to start a guided refinement tour~\cite{meh2017b} through the semantic space. This targeted refinement zooms the canvas on the most  uncertain part of the concept space and suggests a refinement that the user can accept or reject to go to the next suggestion. 

\section{Interactive Learning of the User's Semantics} % Refinement/
\label{sec:ml}
 The foundations of our mixed-initiative technique are the user guidance and learning components. 
These constantly monitor the quality of the concept space to tailor the suggestions for refinement of the most uncertain areas. However, as \textit{serendipitous} exploration has been deemed useful for content exploration~\cite{Serendip2014}, our approach is designed to encourage exploratory refinement. When needed, users can request\textit{ guided refinement suggestions} on demand.  Both of these options are part of the \textit{concept refinement}, designed to teach the system the users' semantics. The counterpart to this knowledge externalization consists of the topic model learning the new semantics, as well as the corresponding adaption of the topics to the learned word associations.  

The main challenge of such an approach is the lack of specificity in the user interactions, i.e., the performed semantic interactions are not directly linked to actionable steps for topic model refinement. % to the semantic relations of keywords in a Corpus to Refine the Topic Modeling... 
We therefore rely on learning the `importance' of words for the given corpus, as well as their relations. 
Hence, the weighted word vectors are the common ground used for learning. 
In addition to corpus-specific concept refinements, we learn global word associations to enable knowledge transfer across comparable document collections. For example, if users refine a concept space for a specific presidential debate, they can reuse that space to initialize the concept extraction for another presidential debate, avoiding a cold start to the second analysis. Overall, for every word in the system we keep track of several scores, including its relevance for every concept, topic, and document, as well as for the corpus and globally. To learn the importance of a word, its level in the concept hierarchy is weighed-in, with super concepts having the largest impact.

During the refinement process, users have two controls to start a new cycle, on-demand. They can update the spatialization of objects by clicking the `\textit{update t-SNE}'-button. On the other hand,  users can retrain the topic modeling by clicking the `\textit{update TM}'-button. To avoid confusion, the positioning of objects on the screen only changes when triggered by the users though these controls.

\subsection{Concept Refinement}
\label{sec:refinement}
As shown in \autoref{fig:Refinement}, we offer users two ways to refine the concept space; (1)~\textit{direct manipulation} and (2)~\textit{guided relevance feedback}. %In the following, both methods will be discussed in more detail.   
Tasks performed during concept refinement include: adjusting the concept hierarchy based on the users' domain knowledge; \textit{cleaning up} potential projection errors; resolving \textit{word chaining} issues (i.e., two words linked through their association with a third, polymorphic word); as well as, finding `\textit{hidden}' concepts based on the topic modeling result. 

Actions can be carried out on selected objects in the canvas. All available commands fall under three primitive types: (1)~change of hierarchy level: (a)~promoting, (b)~demoting; (2)~change of a parent-child relationship: (a)~reassign children, (b)~reassign parent; (3)~splitting or merging siblings. Within the concept hierarchy, every level supports certain interactions:  \textit{super concepts} are not interactive; \textit{concepts} support (1b, 2a,  3); \textit{descriptors} support (1, 2, 3); while \textit{base words} support (1a).

\paragraph{Direct Manipulation}
As described in \autoref{sec:concept-view}, users can directly interact with words in the concept view. The interactions available in the context menu change based on the selected object types. Users are typically offered  sophisticated interactions that combine more than one of the three primitive actions. For example, as depicted in Fig.~\autoref{fig:DirectManipulation}, for a selection of descriptors, users can `\textit{create a concept from the selection}.' This is equivalent to (1a)~promoting the descriptor clicked (in this case: `\textit{college}') to become a concept, then (2)~assigning all other descriptors in that selection to the newly created concept.

\paragraph{Guided Relevance Feedback}%Optimization}
Actions can also be recommended to the user. The intuition behind this targeted refinement is that the system offers users a \textit{guided tour} through the data-space, pointing them to potential problems, with the goal of achieving \textit{maximum gain for minimum feedback}.  To start a tour, users click the `\textit{magic wand}'-button~\inlinegraphics{magicwand} which opens up a suggestion window, displaying the first refinement recommendation, as shown in Fig.~\autoref{fig:GuidedInteraction}. Simultaneously, the concept map gets zoomed to the region of refinement, centralizing and highlighting the objects concerned. 
Users can then accept or reject the suggestion, or choose a different interaction to perform.  
Internally, this action prompts the \textit{refinement recommender} to reevaluate the semantic space based on information collected through the constant \textit{quality monitoring} and fill up the recommendation queue.  

\textbf{Quality Monitoring -- } In order to make more informed decisions for the user guidance, the system tracks several quality criteria across actions: the scatter of word clusters (words in the same concept cluster or neighborhood) based on \textit{cluster-density}, \textit{intra-cluster variance}, and \textit{inter-cluster variance}. Here, we rely on several cluster validity measurement techniques~\cite{ivancsy2005analysis}, including the \textit{root-mean-square standard deviation}~\cite{halkidi2002clustering}, as well as the \textit{S\_Dbw validity index}~\cite{halkidi2002clustering}.
In addition, for \textit{every word}, we keep track of its \textit{neighborhood count}, \textit{semantic similarity to its children and/or parent}, as well as its \textit{spatial distance to children and/or parent}. Furthermore, the quality monitoring component evaluates the \textit{internal quality of the topic modeling} based on the criteria outlined in our previous work~\cite{el2018visual}.

\textbf{Refinement Recommendation -- } Based on the results of the quality monitoring, the recommender keeps a constantly-updated queue of words and their suggested actions.  This queue is formed from words that are chosen based on their importance to the corpus using  \textit{tf-idf scoring}~\cite{sparck1972statistical}. The intuition is that users should be presented with refinements affecting  the \textit{worst-performing}, high-impact words, in order to achieve substantial improvements and give \revision{\textit{minimum feedback for maximum gain}}.  Words important to the whole corpus (high tf-idf) should be concepts, while important for single documents (low tf-idf) should be descriptors.
% intuition: (high tf-idf = important to the whole corpus = should be concepts) (low tf-idf = important for single documents = should be descriptor)
After retrieving the top 50 high-impact words using tf-idf, the recommender loop starts by (1)~\textit{ranking} them based on the quality metrics; (2)~\textit{sorting} possible refinement actions for each word, as well as word clusters, using a decision tree; and finally, (3)~\textit{adding} the words along with their most suitable recommendation to the queue.   The queue is reevaluated if the concept space changes. The recommended actions come from the three primitive interaction types listed above. \revision{An example of a recommendation is depicted in \autoref{fig:teaser}.}

\subsection{Topic Modeling Adaptation}

As described in  \autoref{sec:topic-view}, we depict the associations (semantic similarity) of topics and documents to concepts through topic glyphs.

\begin{wrapfigure}[4]{r}{0.15\columnwidth}
 \vspace{-24pt}
  \begin{center}
    \includegraphics[width=0.19\columnwidth, right]{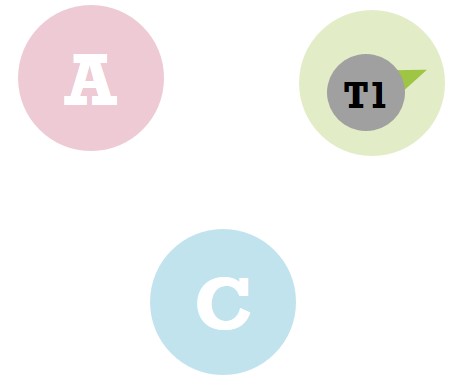}
  \end{center}
%    \vspace{-22pt}
\end{wrapfigure} \noindent During refinement, four different  cases of   topics (and documents) can be observed. \textbf{(1)~\inlinegraphics{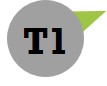}~Single-Concept Topics} are  related to only one concept and placed atop that concept, the corresponding glyph has no large spikes.   

\begin{wrapfigure}[4]{l}{0.15\columnwidth}
 \vspace{-24pt}
  \begin{center}
    \includegraphics[width=0.19\columnwidth, left]{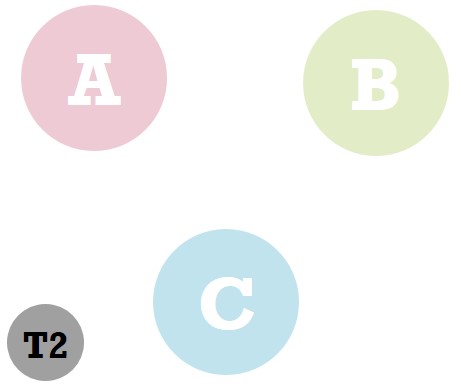}
  \end{center}
%    \vspace{-22pt}
\end{wrapfigure} \noindent  These topics are coherent with the concept structure and do not require further refinement. \textbf{(2)~\inlinegraphics{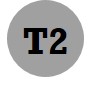}~Unrepresented Topics} that are not related to any concept. They are placed atop an empty region in the concept space, with no large spikes visible in their glyphs. If relevant, users can introduce a respective new concept to the empty area, otherwise, the top topic key-

\begin{wrapfigure}[4]{r}{0.15\columnwidth}
 \vspace{-24pt}
  \begin{center}
    \includegraphics[width=0.19\columnwidth, right]{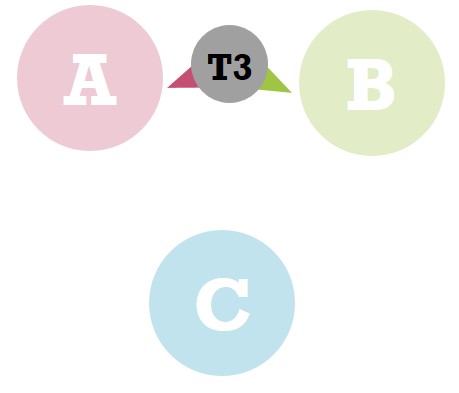}
  \end{center}
%    \vspace{-22pt}
\end{wrapfigure} \noindent  words have to be inspected further. \textbf{(3)~\inlinegraphics{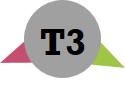}~Multi-Concept Topics} are related to a close neighborhood of concepts. They are placed in-between the related concepts, with small spikes to these concepts in the corresponding glyph.  If closely related, users can merge the concepts, other-

\begin{wrapfigure}[4]{l}{0.15\columnwidth}
 \vspace{-24pt}
  \begin{center}
    \includegraphics[width=0.19\columnwidth, left]{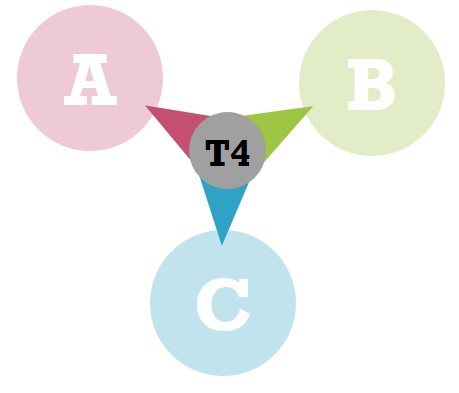}
  \end{center}
%    \vspace{-22pt}
\end{wrapfigure}
\noindent wise, no refinement is needed. \textbf{(4)~\inlinegraphics{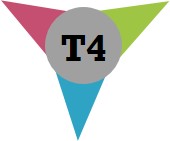}~Concept-Incoherent Topics} are related to multiple concepts \textit{across} the space. They are placed in-between the related concepts according to similarity, with large spikes pointing to these concepts in the corresponding glyph. This is the most critical case, it has to be resolved through targeted concept refinement and a closer inspection of the topic hierarchy.

In the refinement, users can investigate words that are deemed important to a topic but are not relevant for concept distinction. These words are typically less descriptive (in terms of their unique semantic contribution) to a topic than modeled by the algorithm. For example, in a topic model, based on a bag-of-words-representation, (frequent) verbs, adjectives, and adverbs can cause documents to seem similar even though they are not. %This known issue of chaining  
To avoid such chaining effects, we use the learned weights and scores from the concept refinement to readjust the keyword weighting for the topic model training. These act as  \enquote{must-link} and \enquote{cannot-link} constraints~\cite{Andrzejewski2009IncorporatingPriors} to introduce the user-defined notion of relevance to the topic modeling. 
Ideally, a stable and deterministic topic model~\cite{el2018visual} should be used in this process. However, probabilistic models are also applicable but take longer to converge due to \textit{inconsistencies} between their individual runs~\cite{Blei2012ProbabilisticModels}.

\section{Evaluation}

\textbf{Methodology -- } 
We evaluated our approach with in three stages. (1)~To asses the \textbf{usefulness and usability} of our technique, we conducted an expert mixed-initiative study~\cite{isenberg2013} with six participants,  involving two phases of semi-structured interviews, as well as a pair-analytics session~\cite{Kaastra2014}. (2)~Based on the concept space refinements of the study results, we automatically computed the \textbf{topic modeling improvements} across eight quality metrics~\cite{el2018visual}. (3)~To assess the \textbf{perceived quality difference}, we asked four independent annotators to rank the quality of five different concept spaces and their associated topic modeling results.  This section reports the results of all three stages, grouped into \textit{quantitative} and \textit{qualitative} insights.  

\textbf{Dataset and Controls -- }
For our evaluation we sought a dataset with a broadly familiar content, where the expected topic distribution is known. %Additionally, 
% To be consistent and 
To ensure comparability with our previous work on topic modeling refinement~\cite{el2018visual,El-Assady2018ProgressiveFramework}, we chose to use the  second US Presidential Debate between Romney and Obama in 2012, as our corpus for all studies. In this dataset, we treat every speaker utterance as a document.   

\textbf{Participants and Tasks -- }
After conducting a pilot study, we designed the expert user study for a target group of people generally interested in politics. Based on our experience in previous works, we envisioned that scholars in the social sciences and the humanities would fit this profile, we therefore invited two political scientists $Pol_{\{1,2\}}$ and  two linguists $Ling_{\{1,2\}}$ to take part in the study. As a control group, we recruited two computer scientists $CS_{\{1,2\}}$ with no prior knowledge in debate analysis. For the annotation task,  we invited two political scientists $Pol_{\{3,4\}}$ and  two linguists $Ling_{\{3,4\}}$.  
Across the three stages, our goal was to assess the technique's support for the four tasks~\cite{explainer2019} of 
\textbf{[T1]}~{understanding}, \textbf{[T2]}~{diagnosis}, and \textbf{[T3]}~{refinement} of the concept space, as well as, \revision{\textbf{[T4]}~{progressively updating} the topic modeling.}

\subsection{Qualitative Results: Expert Feedback} 
The six sessions of the expert study, 1.5h each, were structured into three parts. We started with a semi-structured interview (40 mins) in which the explanation of the approach was interwoven; we checked the expectations of the participants before introducing new concepts.  Second, we gave the participants full control over the tool and asked them to refine the concept space. In this pair-analytics session (30 mins), we encouraged participants to think-aloud. Lastly, we ended the study with another semi-structured interview (20 mins) that incorporated the participants' expectation statements from the first part, as well as a reflection of the analysis process. All sessions were screen-captured and audio-recorded for further analysis.

\paragraph{Initial Feedback}

Regardless of their prior experience with topic modeling or familiarity with the data, all experts saw benefits in our technique and potential application areas. Some of them were more familiar with automatic content analysis, like $Ling_1$, who stated that she ``\textit{[had] a mixed experience with using topic models, [as] they sometimes extract useful concept but often contain nonsense words.}'' 
On presenting her the idea of our approach she commented: ``\textit{I find it a very helpful concept to be able to include prior knowledge, we often do that after modeling, but it is also good to do before.}'' 
 On the other hand, $CS_1$ declared that he had no previous experience in using topic modeling but has developed similar machine learning techniques before. He observed that ``\textit{the idea of relevance feedback is good because you adapt what you see, but with a topic modeling black-box, you don't know what the machine learns, which [he] would like to be able to evaluate.}'' 

When asked about his expected workflow, $Pol_2$ was quite certain with the way he wanted to proceed in the refinement, stating that he ``\textit{would try to differentiate between situational and general concepts.}'' However, when questioned about the guided refinement, he responded that ``\textit{[he is] not sure about the guidance component, it might speed up the process, but it's not transparent.}'' 
$CS_2$, on the other hand, described his expected workflow as follows: 
``\textit{My workflow would be to start with looking at concepts, to figure out outliers, then group concept descriptors.}'' 

% Case Studies and 
\paragraph{Observations During the Refinement Process}

Before they started their interactive sessions, we asked participants to rate different concept space abstractions. We got mixed feedback on the favored entry point to the analysis from the different participants. $Pol_1$, for example, stated that she would be interested in ``\textit{the system to suggest a full space and [that she] would clean up the unsuitable details}''    As opposed to this \textit{bottom-up refinement}, $Ling_1$ said: ``\textit{A full space is too cluttered for me, I need to understand the problems, then build up my solution.}''  $Ling_2$ also favored a \textit{top-down refinement} stating that ``\textit{starting with fewer concepts is helpful for the exploration.}''  
All participants agreed that interactively choosing an entry point to the analysis is a desirable functionality that they would make use of, depending on their tasks and data, different abstraction levels would make sense for the refinement. 

The participants then generally continued with exploring 
the semantic space first, to get a  better understanding of the corpus. At this stage many %of them 
used the x-ray lens to explore empty regions in the concept and topic views, or find related words to an object. For instance, $Pol_1$ was interested in exploring a topic on \textit{gun violence} and used the lens to find related %documents and 
keywords. After some refinements, she remarked that
``\textit{it is satisfying to clean this mess and see the model respond,}''  describing the interface as a ``\textit{neat combination of ecstatically pleasing components.}''

During the exploration of the space, $Ling_2$ pointed out that ``\textit{the tool is good at identifying communities.}''  
After observing the the concept
regions, $Ling_1$ found a region in the space that she deemed incoherent, commenting: ``\textit{This is a fuzzy area.}'' The contained descriptors included a \textit{moderation cluster}, \textit{temporal keywords}, and \textit{person names}, she pointed out that ``\textit{[she] wouldn't use these words.}'' She continued by selecting the entire region using the lasso tool and deleting its content. She then stated building up new concepts that, in her opinion, described the underlying phenomena more accurately than the previous space. She then updated the projection and topic modeling, observing a better semantic representation. % after refinement.      
The effects of direct manipulation were also praised by $Pol_2$; he stated ``\textit{I can now build my own semantics and theories to test out.}''  Similarly, after observing some positive changes in the semantic space, $Ling_2$ said:  
``\textit{It's like adding my intuition to a stupid machine.}''   
She also commented on the guided refinement, finding the suggested operations useful, only disagreeing with one proposed action.% She, thus, observed that ``\textit{all recommendations worked well.}'' %--  --> $Ling_2$

Overall, the refinement process was well received by all users. Most of them went through several iteration cycles (up to eight during 30 mins), often trying out the effects a refinement would have on the topic modeling and claiming that they had a better intuition of the expected model after the first few cycles. When asked about their final goal for refinement (or stopping criterion), most participants stated that it would be a trade-off between  the importance of the result (e.g., when used for further analysis) and their familiarity with the domain semantics. 
$Pol_2$ observed that ``\textit{not every topic has to be coherent to be helpful,}'' meaning that his final goal would not be to make each topic perfectly fit only one concept but rather to make them meaningful.

\paragraph{Usability and General Assessment}
All experts enjoyed interactive refinement session. In her general feedback, $Ling_1$ immediately cautioned that 
``\textit{it's so easy to use topic modeling results in a wrong way, I find it good to explore the space and understand the reasons for the results.}'' The same sentiment was shared by $Ling_2$, who stated:
``\textit{I like the idea that I can use my knowledge to put things in order, that's really useful and very satisfying.}'' On the other hand, $Pol_2$ proposed to use our approach for communication, saying: ``\textit{I would like to use this tool for presentation, it would be a nice feature to animate though the regions and create a storyline}''  However, he also requested that 
``\textit{[he] would like to be able to track the changes happening in the topic modeling over the different refinement cycles.}'' %Chris: this is not about improvements... When asked about potential improvements $Pol_1$ exclaimed: ``\textit{It just works!}'' He pointed out that the ``\textit{best feature is that this tool provides an overview, so [she] can finally understand the big picture.}'' 

Reflecting his workflow $CS_2$ commented: ``\textit{I was adjusting concepts to reduce topic  spikes, but then I started asking why a topic is [placed] there and how the system understands my interactions.}'' He continued that ``\textit{the color encoding of the words was useful to find outliers.}'' However, $CS_1$ commented that
``\textit{since [he is] not an expert in with these data, [he]  would like to verify the performance of the topic modeling automatically.}'' Nevertheless, ``\textit{[he] like[d] the design of the interactions, [that he] can insert words directly on the canvas and not in a side panel.}''

The most notable additional features suggested by experts include; a feature to start typing to autocomplete a concept and jump anywhere in the space ($CS_2$); to add a space-out or blow-up button for a selected area ($Ling_1$); to ``deep-search'' for similar concepts ($Ling_1$), and finally to enable zoom-dependant resizing of text labels ($Pol_2$).

\subsection{Quantitative Results: Quality Assessments}
Based on the logging results of the expert studies, we can compare the automatic quality metrics for the topic modeling.
We did not observe any significant difference in the refinement results across the three user groups. Experts and non experts, alike, were able to enhance the topic modeling results through our technique.  
The average relative change, from the initial model to the refined model, based on the eight observed quality metrics~\cite{el2018visual} was as follows: 
\textbf{Coherence}~(-5.49\%);
\textbf{Separation}~(-12.09\%);
\textbf{Distinctiveness}~(331.31\%);
\textbf{Point-wise Mutual Information}~(4.32\%);
\textbf{Certainty}~(0.66\%);
\textbf{Branching Factor}~(-26.47\%);
\textbf{Compactness}~(-11.77\%);
and \textbf{Topic Size}~(1.45\%).
Hence, while on average the topics became slightly less coherent and separated, they became \textit{significantly more distinct} during the refinement. 
                                   
% after using the concept space to improve the topic modeling

The last stage of our evaluation is the assessment and ranking of five models by four annotators. The manual refinement model was generated by a participant in the first study. The other models were created following the guided refinement suggestions, the initial model at the default level of abstraction, and the initial model at high and low abstraction levels. Annotators were given the agenda of the debate and asked to rank the concept view, as well as a keyword-list of the corresponding top topic descriptors. In our annotation guidelines, we asked them to base their ranking on four criteria: completeness, coherence, separability,  descriptiveness. The results, as shown in \autoref{tab:annotation-result-overview}, confirm that the manual refinement of the concept space yields the most well-perceived concept view, while the guided topic refinement leads to the highest ranking topic modeling result. This might be due to overlooked, uncertain regions during the manual refinement. In their annotations, they pointed out that incoherent areas are mainly comprised of clusters of names or  general words, like \textit{tomorrow}, \textit{country}, etc.  This suggests a potential improvement through domain-specific filters for non-informative words.

% \textbf{Completeness, Coherence, Separability,  Descriptiveness}

% They pointed out name clusters and general words like tomorrow, country, etc.
% --> desire to have a domain-specific filter for non-informative words

% Some didn't like verbs others did Same with country names
% --> desire to weight categories of words in the first preprocessing step

\begin{table}
  \centering
  \vspace*{2mm}
  \resizebox{\linewidth}{!}{%
  \begin{tabular}{c|cc} \toprule
  Compared Output & Semantic Concept Space  & Topic Modeling Result\\ 
  \midrule
\textbf{Manual Refinement} &	\textbf{1.00}	(0.00)	& 2.00	(0.70) \\
\textbf{Guided Refinement} &	2.25 (0.43)		& \textbf{ 1.50} (0.51) \\
\textcolor{gray}{\textbf{Default Model}} & \textcolor{gray}{3.25	(0.82)}		& \textcolor{gray}{2.75	(1.09)}  \\
\textbf{High-Abstraction} &	4.00	(0.71)		& 4.25	(0.82) \\
\textbf{Low-Abstraction} &	4.50	(0.87)		& 4.50	(0.50) \\
  \bottomrule
  \end{tabular} }  
  \vspace*{-1mm}
  \caption{Ranked output of the concept space and the corresponding topic modeling %result  
  (scale: 1--best, 5--worst) according to the annotators' perception of quality, the standard deviation is shown in parentheses.}
  \label{tab:annotation-result-overview}
  \vspace*{-15pt}
\end{table}

% \newpage
\section{Discussion and Conclusion}

Typically statistical machine learning algorithms, such as topic models, do not incorporate the semantic relations between objects. It is rare that topic models consider the semantic similarity between keywords or documents. Rather, they rely on keyword scoring and statistically induced relations between objects to group them. 
On the other hand, domain experts see implicit relations between objects and attributes that they cannot incorporate into the machine learning models. Modeling user semantics independent of the topic model is challenging as the user model cannot be tuned to give the best topic model outputs without rapid, iterative feedback. Typically feedback is constrained by algorithmic parameters and is
% , therefore, 
challenging for non-machine-learning-experts~\cite{El-Assady2018ProgressiveFramework}.

Semantic Concept Spaces\footnote{The system is available as part of the  \href{http://lingvis.io/}{lingvis.io} Framework~\cite{lingvisio} under: \href{http://concept-spaces.lingvis.io/}{http://concept-spaces.lingvis.io/}}$^,$\footnote{This work has received funding from the DFG/SPP-1999 VALIDA project (number 376714276) and the DFG Research Unit FOR2111/QI project 8. It was further supported by the SFB/Transregio 161 (number 251654672), projects A03 and A04, as well as NSERC Canada Research Chairs.} 
contributes an approach to bring the expression of user semantics \textit{closer} to the actual ML model. Users can see the \textit{big picture} of the concept space and generate new ground truth data through their interactions and knowledge externalization, a form of ``machine teaching''. 
The modifications applied to the concept space become transferable knowledge which can be used to initiate models used on other data. 
% Throughout this process, the specific ML model (topic model, in this work) is not visible to the end user. 
We introduce a design that is based on two parallel hierarchies: the concept and the topic hierarchies.  The interactions take place in this data space and thus the whole system is ``model agnostic.''

In the future we envision improvements to the recommendation system, for example, verbalizations of system decisions~\cite{SBE+18a} could help a user know the reasoning behind a recommended interaction and provide  better guidance for model refinement. 
Because the goal of this project is to capture the semantics of natural language, there is also a lot of opportunities to engage with this data through a natural language interface, for example, verbally expressing a list of descriptors which would make a concept  faster than searching the concept space and adding them. 
Study participants expressed a desire to compare consecutive topic models during the refinement process. This could potentially be achieved through topic matching~\cite{Assady18LTMA}.

\bibliographystyle{abbrv-doi}
% \bibliography{sample-bibliography}
\bibliography{Bibliography}
\end{document}